%% file: main.tex
\documentclass[sigconf]{acmart}
\AtBeginDocument{%
  }

\copyrightyear{2026}
\acmYear{2026}
\setcopyright{cc}
\setcctype{by}
\acmConference[WWW '26]{Proceedings of the ACM Web Conference 2026}{April 13--17, 2026}{Dubai, United Arab Emirates}
\acmBooktitle{Proceedings of the ACM Web Conference 2026 (WWW '26), April 13--17, 2026, Dubai, United Arab Emirates}
\acmPrice{}
\acmDOI{10.1145/3774904.3792670}
\acmISBN{979-8-4007-2307-0/2026/04}

\usepackage[utf8]{inputenc} 
\usepackage[T1]{fontenc}    
\usepackage{hyperref}       
\usepackage{url}            
\usepackage{booktabs}       
\usepackage{amsfonts}       
\usepackage{nicefrac}       
\usepackage{microtype}      
\usepackage{xcolor}         

\usepackage{graphicx}
\usepackage{caption}
\usepackage{subcaption}
\usepackage{bm}
\usepackage{tabularx}
\usepackage{multirow}
\usepackage{makecell}
\usepackage{xr}
\usepackage{placeins}
\usepackage{enumitem}
\usepackage{array}
\usepackage{float}
\usepackage{dblfloatfix}
\usepackage{makecell}
\usepackage{amsmath}
\usepackage{algorithm}
\usepackage{algpseudocode}
\usepackage{tikz}
\usetikzlibrary{positioning, arrows.meta}

\newcolumntype{C}[1]{>{\centering\arraybackslash}m{#1}}
\fontsize{8.9}{9}\selectfont

\begin{document}

\title{Post-hoc Popularity Bias Correction \\in GNN-based Collaborative Filtering}


\author{Md Aminul Islam}
\affiliation{%
  \institution{University of Illinois Chicago}
  \city{Chicago}
  \state{IL}
  \country{USA}
 }
\email{mislam34@uic.edu}

\author{Elena Zheleva}
\affiliation{%
  \institution{University of Illinois Chicago}
  \city{Chicago}
  \state{IL}
  \country{USA}
 }
\email{ezheleva@uic.edu}

\author{Ren Wang}
\affiliation{%
  \institution{Illinois Institute of Technology}
  \city{Chicago}
  \state{IL}
  \country{USA}
}
\email{rwang74@illinoistech.edu}

\renewcommand{\shortauthors}{Md Aminul Islam, Elena Zheleva, \& Ren Wang}

\begin{abstract}
    \input{0-abstract}    
\end{abstract}



\maketitle
\newcommand\webconfavailabilityurl{https://doi.org/10.5281/zenodo.18371801}
\ifdefempty{\webconfavailabilityurl}{}{
\begingroup\small\noindent\raggedright\textbf{Resource Availability:}\\
The source code of this paper has been made publicly available at \url{\webconfavailabilityurl}.
\endgroup
}

\input{1-introduction}
\input{2-related-work}
\input{3-preliminaries}
\input{4-methodology}
\input{5-experimental-setup}
\input{6-results}
\input{7-conclusion}

\section*{Acknowledgments}
This research was partly supported by the National Science Foundation under Award No. 2217023.

\bibliographystyle{ACM-Reference-Format}
\balance
\bibliography{main}

\appendix
\input{8-appendix}

\end{document}

%% file: 0-abstract.tex
User historical interaction data is the primary signal for learning user preferences in collaborative filtering (CF). However, the training data often exhibits a long-tailed distribution, where only a few items have the majority of interactions. CF models trained directly on such imbalanced data are prone to learning popularity bias, which reduces personalization and leads to suboptimal recommendation quality. Graph Neural Networks (GNNs), while effective for CF due to their message passing mechanism, can further propagate and amplify popularity bias through their aggregation process. Existing approaches typically address popularity bias by modifying training objectives but fail to directly counteract the bias propagated during GNN's neighborhood aggregation. Applying weights to interactions during aggregation can help alleviate this problem, yet it risks distorting model learning due to unstable node representations in the early stages of training. In this paper, we propose a \textit{Post-hoc Popularity Debiasing} (PPD) method that corrects for popularity bias in GNN-based CF and operates directly on pre-trained embeddings without requiring retraining. By estimating interaction-level popularity and removing popularity components from node representations via a popularity direction vector, PPD reduces bias while preserving user preferences. Experimental results show that our method outperforms state-of-the-art approaches for popularity bias correction in GNN-based CF.

%% file: 1-introduction.tex
\section{Introduction}
Recommender systems are integral to modern online platforms, shaping user experiences through personalized recommendations across various domains, including e-commerce, entertainment, and social networking. By leveraging historical user–item interactions, they present users with relevant content from large item collections, a process commonly referred to as collaborative filtering (CF)~\cite{he-www17, su-aai09}. Recently, Graph Neural Networks (GNNs) have emerged as a powerful method for CF by modeling recommendations as a user–item bipartite graph and propagating information to multi-hop neighborhoods via neighborhood aggregation~\cite{yang-sigir21}. The aggregation mechanism enables GNN-based CF methods to learn higher-order interaction patterns, resulting in high-quality user and item representations and state-of-the-art performances~\cite{he-sigir20, gao-www22}.

\begin{figure} [hb]
  \centering
  \captionsetup{justification=raggedright, margin=0cm}
  \includegraphics[width=0.32\textwidth]{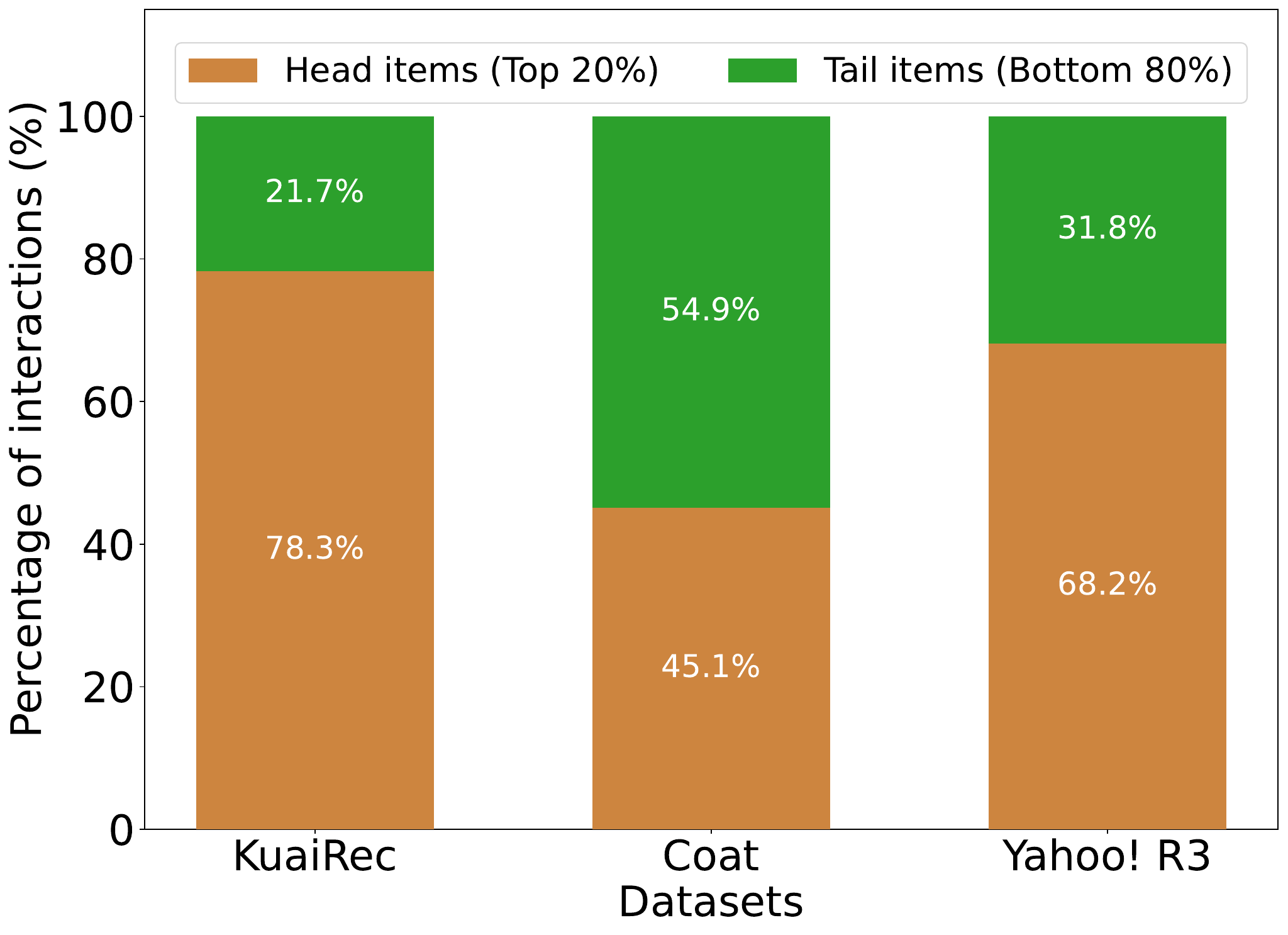}

  \captionsetup{width=0.48\textwidth}
  \caption{Distribution of interactions across head (top \bm{$20\%$}) and tail (bottom \bm{$80\%$}) items in the KuaiRec, Coat, and Yahoo! R3 datasets. Most interactions are concentrated in head items, highlighting the data imbalance  in these datasets.}
  \label{fig:pop_bias_demonstration}
\end{figure}

Recommender systems often rely on user interaction data, such as clicks, to infer preferences, since this data can easily be collected from interaction logs. However, such data is prone to biases~\cite{ai-sigir18, joachims-wsdm17}, with popularity bias being particularly severe~\cite{zhou-sigir23}. Interaction data in recommender systems is often long-tailed, with a small set of items receiving most feedback while the majority remain rarely interacted with~\cite{zhang-neurips22}. 
CF models trained on such skewed data inherit popularity bias~\cite{yao-neurips17, canamares-sigir18}, over-recommending head items and under-recommending tail items. Through the feedback loop in recommender systems~\cite{chaney-recsys18}, this bias can be amplified over time, reducing personalization and weakening long-term performance. Figure~\ref{fig:pop_bias_demonstration} illustrates the effect of data imbalance in interaction data.
In the KuaiRec, Coat, and Yahoo! R3 datasets, items are sorted by interaction frequency, with the top $20\%$ categorized as head items and the remaining $80\%$ as tail items. User interactions are disproportionately concentrated on head items. 
Consequently, training models on such skewed data amplifies head-item recommendations, further marginalizing tail items in future recommendations.

Several approaches have been proposed to mitigate popularity bias. A common strategy is to adjust the training loss to reduce the influence of popular items. Inverse propensity weighting (IPW) methods~\cite{gruson-wsdm19, joachims-wsdm17, ai-sigir18} reweight the loss by upweighting unpopular items and downweighting popular ones, while regularization-based methods~\cite{boratto-ipm21, wasilewski-flairs16, chen-sigir20} penalize correlations between predicted scores and item popularity. Causal graph-based methods~\cite{ning-www24, wei-sigkdd21, bonner-recsys18, wang-kdd21} instead use causal structures to identify sources of popularity and apply causal inference 
techniques to mitigate their effect on relevance estimation. However, all three of these loss-based approaches may not fully eliminate popularity bias in GNN-based CF~\cite{zhou-sigir23}.

While GNNs are highly effective for CF tasks, they tend to amplify popularity bias~\cite{chen-front24, zhou-sigir23}. Due to the long-tailed nature of interaction data, popular items connect to many more users than unpopular (niche) items, giving them disproportionate influence during propagation~\cite{zhou-sigir23}. During the message passing process, high-degree nodes (i.e., popular items) propagate their information more quickly to neighboring users than low-degree nodes (i.e., niche items), causing users’ representations to move closer to these popular items in the embedding space~\cite{chen-front24, chen-aaai20}. As a result, learned representations tend to be dominated by signals from high-degree items, and repeated propagation across multiple layers spreads this effect to higher-order neighbors, further strengthening the dominance of popular items~\cite{chen-front24}. Consequently, popular items become similar to a wide range of users in the latent space, making them more likely to receive higher prediction scores~\cite{zhou-sigir23} and appear frequently in many users’ recommendations.

To address this, recent works~\cite{kim-cikm22, zhou-sigir23} weight interactions during aggregation, akin to propensity scores, to downweight popular item interactions during aggregation. However, these methods assign weights in every training epoch—either using pre-computed scores~\cite{kim-cikm22} or dynamically estimated values~\cite{zhou-sigir23}. Since model representations are unstable in the early stages, the resulting embeddings are under-trained and noisy, which leads to unreliable popularity estimates that can distort and mislead the learning process.

Both loss-based and aggregation-weight methods require retraining the backbone, making them impractical for deployed models~\cite{chen-aaai20}. Post-processing re-ranking methods~\cite{abdollahpouri-arXiv19, zhu-kdd21} adjust predicted scores to diversity or popularity bias but cannot remove popularity effect embedded in node representations. More recently, a post-hoc method~\cite{chen-front24} for GNN-based CF estimates popularity effects using simple node degrees, without considering personalized preferences or interaction-level popularity effects. One might assume that debiasing can be achieved by simply removing some popular items from the recommendations to match the true underlying distribution. This distribution, however, is unknown in practice. Moreover, since popularity is not always harmful~\cite{zhao-kde22}, indiscriminately suppressing head items risks discarding truly relevant items of user interest.


In this paper, we propose a \textit{Post-hoc Popularity Debiasing} (PPD) method to mitigate popularity bias in GNN-based CF. Our method operates on pre-trained embeddings from an already trained model. We estimate the popularity of each user-item interaction by combining global preference with personalized preference derived from the embeddings. Based on these estimates, we identify a popularity direction vector for each node. We then project the node embedding onto this direction to obtain its component along the direction and update the embedding by removing this popularity component. This eliminates the popularity effect while preserving preference-driven signals, resulting in unbiased node representations for generating unbiased recommendations. Though this general idea is simple, this method provides a post-hoc solution that is very effective at reducing popularity bias without retraining the model.

Our method has several advantages over existing popularity bias correction methods. Unlike loss-based methods~\cite{boratto-ipm21, wasilewski-flairs16, gruson-wsdm19, ning-www24, wei-sigkdd21, wang-kdd21}, it requires no loss modification, making it applicable to any GNN-based CF model. Second, it operates in a post-hoc manner, eliminating the need for retraining and allowing application to already deployed models. Third, instead of weighting interactions in aggregation during training~\cite{kim-cikm22, zhou-sigir23}, our method operates directly on learned embeddings, allowing popularity estimation on stable representations. Finally, unlike prior GNN-based post-hoc method~\cite{chen-front24}, it estimates popularity at the interaction level for finer-grained effects and captures both global and personalized influences in popularity estimation, rather than relying on simple node degrees.
\vspace{-0.3cm}

%% file: 2-related-work.tex
\section{Related Work}
Recommender systems often exhibit popularity bias, causing popular items to dominate recommendations~\cite{chen-sigir22, zhang-neurips22, chen-sigir20, rhee-recsys22}. Existing methods to address this include IPW, post-processing re-ranking, regularization, and causal inference methods.

IPW-based methods mitigate popularity bias by down-weighting frequent items, assigning each interaction a weight inversely proportional to the item's popularity~\cite{ai-sigir18, joachims-wsdm17}. To reduce the high variance of IPW, subsequent works~\cite{gruson-wsdm19, bottou-jmlr13} use normalization or smoothing penalties. However, the effectiveness of IPW depends on accurate propensity score estimation~\cite{ovaisi-sigir21}. IPW is also widely used in learning-to-rank systems to correct biases such as position and selection bias~\cite{joachims-wsdm17, oosterhuis-sigir20, wang-sigir16, ai-sigir18, luo-sigir24, luo-wsdm23}. Regularization-based methods incorporate constraints during training to reduce popularity. LapDQ~\cite{wasilewski-flairs16} promotes diversity via item-distance regularization, INRS~\cite{kamishima-recsys14} applies mean-matching to favor less popular items, ESAM~\cite{chen-sigir20} transfers knowledge from popular to unpopular items to alleviate sparsity, and Sam-Reg~\cite{boratto-ipm21} reduces the correlation between relevance and popularity. Post-processing re-ranking methods operate after recommendation generation, leaving user and item representations intact. They address goals such as preserving historical preference distributions (calibration)~\cite{steck-recsys18}, enhancing diversity~\cite{abdollahpouri-arXiv19}, and correcting popularity bias in dynamic recommender systems by adjusting predicted scores~\cite{zhu-kdd21}. 
Causal inference has been used to separate user interests from popularity effects~\cite{bonner-recsys18, wang-kdd21, wei-sigkdd21, ning-www24, he-icds22, zhao-kde22}. CausE~\cite{bonner-recsys18} uses causal graphs for unbiased learning, MPCI~\cite{he-icds22} blocks paths from popularity to predictions, MACR~\cite{wei-sigkdd21} removes direct popularity effects via multi-task counterfactual learning, and PPAC~\cite{ning-www24} jointly models global and personal popularity within user neighborhoods using counterfactual inference.

Deep learning-based recommender systems rely on high-quality representations to predict user–item interactions~\cite{zha-kdd22}. GNNs achieve this by aggregating information from neighboring nodes~\cite{chen-sigir22, chen-cikm20, dong-www23, huang-icml22, zhang-wsdm23, zhang-sigir22}, with propagation layers capturing higher-order connectivities. They are state-of-the-art for CF due to this ability. LightGCN~\cite{he-sigir20} simplifies GNNs by removing non-linear transformations and activations and is widely used in GNN-based CF. NIA-GCN~\cite{sun-sigir2020} models pairwise neighbor relationships, while UltraGCN~\cite{mao-cikm21} addresses over-smoothing by approximating infinite-layer limits. With the rise of self-supervised learning, methods like SGL~\cite{wu-sigir21} and NCL~\cite{lin-www22} enhance robustness by combining contrastive objectives with structural and semantic signals, while DirectAU~\cite{wang-kdd22} emphasizes alignment and uniformity in representations. SimGCL~\cite{yu-sigir22} and XSimGCL~\cite{yu-kde23} further show that contrastive learning improves recommendations by promoting uniform user–item embeddings.
However, these approaches overlook popularity amplification in GNNs. APDA~\cite{zhou-sigir23} addresses this by reweighting interactions using inverse popularity scores during aggregation, and NAVIP~\cite{kim-cikm22} applies IPW during aggregation based on interaction counts. Apart from these, Zhang et al.~\cite{zhang-kdd23} propose a Cross Decoupling Network (CDN) that decouples memorization and user distributions to mitigate long-tail bias in content-based recommender systems. Recent studies have explored out-of-distribution (OOD) generalization in GNN-based recommendation. DRO~\cite{zhao-www25} mitigates the effect of noisy samples, CausalDiffRec~\cite{zhao-www25(1)} removes environmental confounders via causal diffusion, and AdvInfoNCE~\cite{zhang-neurips23} improves generalization using a hardness-aware adversarial contrastive loss. Despite their effectiveness, these methods do not explicitly address popularity bias. DAP~\cite{chen-front24} is a post-hoc debiasing approach that estimates popularity via clustering similar nodes and leveraging signals from high- and low-degree neighbors, and then removes popularity information from node embeddings.

%% file: 3-preliminaries.tex
\section{Problem Description}
\subsection{Preliminaries}
Let $\mathcal{U} = \{u_1, u_2, \ldots, u_M\}$ be the set of users, and $\mathcal{I} = \{i_1, i_2, \ldots, i_N\}$ be the set of items. For each user-item pair $(u, i)$, we define a binary variable $y_{ui}$, where $y_{ui}=1$ indicates that user $u$ has interacted with item $i$, and $y_{ui}=0$ otherwise. Based on these interactions, we can construct a bipartite graph $\mathcal{G}=(\mathcal{V}, \mathcal{E})$, where the node set $\mathcal{V}$ contains both users and items, and the edge set $\mathcal{E}$ consists of pairs $(u, i)$ for which $y_{ui}=1$. The goal is to recommend the top-$k$ items that a user $u$ has not interacted with but is most likely to interact.

In recent years, Graph Convolutional Networks (GCNs) have become powerful for learning representations of users and items in recommender systems~\cite{mao-cikm21, kipf-iclr17}. GNN-based CF methods~\cite{he-sigir20, wang-sigir19, wang-sigir20, wu-sigir21, lin-www22, yu-sigir22, yu-kde23} aggregate information from neighbors. One graph convolution block of GNN-based CF methods can be written as:
\begin{equation} \label{eq:a_node_prop}
    \begin{aligned}
        \mathbf{e}_u^{(l)} = f_{\text{propagate}}\Big(\{\mathbf{e}_i^{(l-1)} \mid i \in \mathcal{N}_u \cup \{u\} \}\Big), \\ 
        \mathbf{e}_i^{(l)} = f_{\text{propagate}}\Big(\{\mathbf{e}_u^{(l-1)} \mid u \in \mathcal{N}_i \cup \{i\} \}\Big), 
    \end{aligned}
\end{equation}
where $\mathbf{e}_u^{(l)}$ ($\mathbf{e}_i^{(l)}$) denotes the embedding of user $u$ (item $i$) at the $l$-th propagation layer and $\mathcal{N}_u$ ($\mathcal{N}_i$) is the set of neighbors of user $u$ (item $i$). The propagation function $f_{\text{propagate}}(\cdot)$ updates user $u$'s (item $i$'s) representation by aggregating information from its neighbors’ $(l-1)$-th layer embeddings, yielding the new embedding  $\mathbf{e}_u^{(l)}$ ($\mathbf{e}_i^{(l)}$) at $l$-th layer. The initial embeddings $\mathbf{e}^{(0)}$ can be an ID-based user and item representation and are learnable parameters associated with each node. For GNN-based CF, the most popular and widely used backbone is LightGCN~\cite{he-sigir20}, which simplifies the message passing mechanism of GCNs while preserving recommendation effectiveness. The propagation function $f_{\text{propagate}}(\cdot)$ in a LightGCN layer is defined as: 
$\mathbf{e}_u^{(l)} = \sum_{i \in \mathcal{N}_u} \frac{1}{\sqrt{d_u d_i}} \mathbf{e}_i^{(l-1)}, \mathbf{e}_i^{(l)} = \sum_{u \in \mathcal{N}_i} \frac{1}{\sqrt{d_i d_u}} \mathbf{e}_u^{(l-1)},$ where $d_u$ $(d_i)$ denote the degrees of user $u$ (item $i$) in the graph. LightGCN preserves the neighborhood aggregation of GCNs while removing extra steps such as feature transformation and non-linear activation,  and uses simple linear embedding propagation.

After $l$ iterations of propagation, the representation $\mathbf{e}_u^{(l)}$ ($\mathbf{e}_i^{(l)}$) encodes information from $l$-hop neighbors. After stacking $L$ propagation layers, the readout function $f_{\text{readout}}(\cdot)$ then combines the representations from all layers to produce the final embedding:
\begin{equation} \label{eq:a_node_fianl_emb}
    \begin{aligned}
        \mathbf{e}_u = f_{\text{readout}}\Big([\mathbf{e}_u^{(0)}, \mathbf{e}_u^{(1)}, \dots, \mathbf{e}_u^{(L)}]\Big), \\
        \mathbf{e}_i = f_{\text{readout}}\Big([\mathbf{e}_i^{(0)}, \mathbf{e}_i^{(1)}, \dots, \mathbf{e}_i^{(L)}]\Big).
    \end{aligned}
\end{equation}
For example, the $f_{\text{readout}}(\cdot)$ function in LightGCN is a mean-pooling operation that combines information from all layers to form the final representation of each node:
$\mathbf{e}_u = \frac{1}{L+1} \sum_{l=0}^{L} \mathbf{e}_u^{(l)}, \mathbf{e}_i = \frac{1}{L+1} \sum_{l=0}^{L} \mathbf{e}_i^{(l)}.$
The predicted score between a user $u$ and an item $i$ is computed as the inner product of user and item final representations, i.e., ${\hat{y}}_{ui} = \mathbf{e}_u^\top \mathbf{e}_i$. Using these scores, the top-$k$ items that user $u$ has not interacted with can be recommended.

The most commonly used objective function in GNN-based CF (e.g., LightGCN) is the pairwise Bayesian Personalized Ranking (BPR) loss~\cite{rendle-arxiv12}, which maximizes the difference between positive and negative item scores with an $L_2$ regularization term:
\begin{equation}
    \mathcal{L}_{BPR} = \sum_{(u,i) \in \mathcal{N}_u, j \notin \mathcal{N}_u} - \ln \sigma(\hat{y}_{ui} - \hat{y}_{uj}) + \gamma \|\Theta\|_2^2.
\end{equation}
Here, $j$ is an item not interacted with by $u$ and $\sigma$ is the sigmoid activation function. $\Theta$ denotes the current batch embeddings of users and items, and $\gamma$ controls the strength of the $L_2$ penalty. Since our method operates on pre-trained embeddings in a post-hoc manner without changing the training objective, it can be adapted to other GNN-based CF models that use different training objectives. For example, our method can also be applied to GNN-based CF models, such as contrastive learning approaches that combine a contrastive loss with BPR 
loss~\cite{lin-www22, wu-sigir21, wu-kdd23, yu-sigir22}, or Neural Graph Collaborative Filtering (NGCF)~\cite{wang-sigir19}, which employs BPR loss with a non-linear activation function in aggregation.

\subsection{Problem statement}
Given user embeddings ($\mathbf{e}_u$) and item embeddings ($\mathbf{e}_i$) learned from a pre-trained model, the objective is to find a transformation \( T \) from a set of possible transformations \( \mathcal{T} \) that best maps the learned embeddings closer to their true unbiased counterparts, $\mathbf{e}_u^{({*})}$ or $\mathbf{e}_i^{({*})}$. In other words, the goal is to identify the transformation that, when applied to the learned embeddings, minimizes their deviation from the true unbiased representations. This can be expressed as:
\begin{equation}
    T^{\star} = 
    \arg\min_{T \in \mathcal{T}} 
    \sum_{j=1}^{|\mathcal{V}|} 
    \left\|
        T(\mathbf{e}_j) - \mathbf{e}_j^{({*})}
    \right\|_2^2.
    \label{eq:problem_def}
\end{equation}

However, in practice, the true unbiased embeddings are not known, so the objective in equation~\eqref{eq:problem_def} cannot be optimized directly. Instead, the goal is to update the learned embeddings by removing popularity-related components from the node representations, so that the updated embeddings more closely resemble the users’ true preference space. The effectiveness of this adjustment can be indirectly evaluated through improvements in predictive accuracy through an unbiased evaluation.

%% file: 4-methodology.tex
\section{Methodology}
In this section, we introduce our PPD method for mitigating popularity bias in GNN-based CF. We first estimate the popularity for each user–item interaction by proposing a novel metric of popularity that quantifies how much the interaction is influenced by item global preference rather than user preferences, and use these estimates to define a popularity direction in the embedding space. We then project node embeddings onto this direction to get the component of node embeddings on this direction, and update the node embeddings by removing this popularity component from the node embeddings. Despite its simplicity, this approach effectively reduces popularity bias in a 
post-hoc manner, without requiring retraining, while preserving the underlying preference signals.

\subsection{Popularity estimation}
User-item interactions can arise from two main factors, namely the global preference of the item and the user’s personalized preference for it. To quantify the contribution of popularity to each interaction, we define three measures: global preference, personalized preference, and popularity score. 

\subsubsection{Global preference} Global preference reflects the fact that some items receive disproportionately high recommendations across the user base. To capture this effect in the embedding space, we measure how similar an item is, on average, to all users. Items with higher similarity with users tend to obtain higher prediction scores and are more likely to be recommended, reflecting the extent to which interactions are driven by broader popularity.

\textsc{Definition 1 (global preference).} \textit{Let $p_i$ be the global preference for item $i$, which estimates its popularity in the embedding space by measuring its similarity to all users, and is defined as:}
\begin{equation} \label{eq:global_popularity_rule}
    p_i = \frac{1}{|\mathcal{U}|} \sum_{u \in \mathcal{U}} {sim}(\mathbf{e}_u, \mathbf{e}_i),
\end{equation}
\textit{where $\text{sim}(\cdot)$ is a similarity function (e.g., dot product or cosine similarity).}

Existing works~\cite{zhang-sigir21, wei-sigkdd21, ning-www24} typically measure an item's global preference by its interaction frequency in the training data, i.e., the ratio of total interactions with the item to the total number of users. In contrast, we define global preference in the embedding space by measuring how similar an item is to users, reflecting how the pre-trained model would tend to promote certain items broadly and thereby makes them popular through recommendations.

\subsubsection{Personalized preference} Personalized preference reflects how well an item matches the historical behavior of a  user. If an interacted item is similar to a user’s historical interactions, it likely reflects the user’s personalized preferences. However, since popular items may appear in a user’s history, we adjust their influence when computing personalized preferences.

\textsc{Definition 2 (Personalized Preference).} \textit{The personalized preference $r_{ui}$ for user $u$ measures how closely item $i$ aligns with the other items interacted with by user $u$, and is defined as:}
\begin{equation} \label{eq:personalized_pref_rule}
    r_{ui} = \frac{1}{|N_u|} \sum_{j \in N_u} \textit{sim}(\mathbf{e}_i, \mathbf{e}_j) - \beta p_i p_j.
\end{equation}
Intuitively, $r_{ui}$ measures how well item $i$ aligns with the historical preferences of user $u$. However, if item $i$ is globally popular and the set $N_u$ also contains other popular items, their embeddings may appear similar in the embedding space.
Consequently, the personalized preference $r_{ui}$ may be high for a pair of popular items. To address this, we introduce a penalty term controlled by the hyperparameter $\beta \geq 0$. The hyperparameter $\beta$ controls how strongly global preference effects are discounted in personalized preference. Larger values impose stronger penalties on personalized preferences, while smaller values allow item similarity to contribute more. It can be dataset-dependent, based on data skewness and sparsity, and therefore we tune $\beta$ as a hyperparameter.

\subsubsection{Popularity score} We define a popularity score for each interaction to measure the influence of an item’s global preference over the user’s personalized preference. We define this score as the difference between the item’s global and personalized preferences.

\textsc{Definition 3 (Popularity Score).} \textit{Let $b_{ui}$ be the popularity score for an interaction $(u,i)$, where $p_i$ is the global preference of item $i$ and $r_{ui}$ is the user's personalized preference for item $i$. Then, $b_{ui}$ is defined as:}
    \begin{equation} \label{eq:pop_bias_scores}
        b_{ui} = p_i - r_{ui}.
    \end{equation}
The popularity score $b_{ui}$ for each interaction estimates the extent to which an interaction between user $u$ and item $i$ is influenced by popularity rather than user preference. The reason for subtracting personalized preference is that some part of an item’s global preference reflects true user preferences. Considering global preference ($p_i$) directly to address popularity bias can hurt recommendations~\cite{zhao-kde22}, as some popular items can truly be relevant.
By removing personalized preference component, the popularity score captures only the residual effect of popularity, isolating the part of the interaction that is driven purely by popularity. 
Without this, preference-aligned popular items can incorrectly be treated as pure popularity.
A high score indicates that the item is popular but not well-aligned with the user’s interests, suggesting a popularity-driven interaction. A low score implies alignment with the user’s preferences, reflecting a more true preference. To make the two scores $p_i$ and $r_{ui}$ comparable, we can normalize them before computing $b_{ui}$. We compute the popularity score for each interaction using the stable user and item embeddings learned by the pre-trained model. This yields a more accurate score, as the embeddings already capture both popularity and meaningful user interests. This is an advantage of our approach over methods~\cite{zhou-sigir23, kim-cikm22} that estimate popularity during training.

\subsection{Post-hoc popularity bias correction via projection}
In the previous section, we estimated interaction-level popularity scores, providing a fine-grained measure of how much each user-item interaction is influenced by item popularity. Popularity-driven components can dominate user and item representations and further amplify bias through GNN's propagation. Our key idea is to use these interaction-level popularity scores to construct a popularity direction in the embedding space for each node, representing how popularity pulls embeddings away from true preferences. Using this popularity direction, we remove popularity-induced components from node embeddings, thereby preserving the underlying preference signals. This debiasing is performed at the initial embedding layer ($l = 0$) of the 
pre-trained model, and then we propagate information, ensuring that unbiased signals are propagated through all subsequent layers of the GCN.

For each user \( u \), we compute two embedding centroids using the items \( \mathcal{N}(u) \) that the user interacted with: a popularity-based centroid and a preference-based centroid. Because $b_{ui}$ represents the extent to which item $i$ is interacted by user $u$ due to popularity, $(1-b_{ui})$ reflects the degree to which the interaction aligns with user $u$'s personal preferences. Based on this idea, we define the popularity-based centroid ($\bar{\mathbf{e}}_{\text{pop}}^{(0)}(u)$) and the preference-based centroid ($\bar{\mathbf{e}}_{\text{pref}}^{(0)}(u)$) for each user $u$ at the initial layer (layer $0$):
\begin{equation} \label{eq:pop_direction_user}
    \bar{\mathbf{e}}_{\text{pop}}^{(0)}(u) = \frac{\sum_{i \in \mathcal{N}(u)} b_{ui} \cdot \mathbf{e}_i^{(0)}}{\sum_{i \in \mathcal{N}(u)} b_{ui} + \epsilon}, \quad
    \bar{\mathbf{e}}_{\text{pref}}^{(0)}(u) = \frac{\sum_{i \in \mathcal{N}(u)} (1 - b_{ui}) \cdot \mathbf{e}_i^{(0)}}{\sum_{i \in \mathcal{N}(u)} (1 - b_{ui}) + \epsilon}.
\end{equation}
The denominator is a normalizing factor, and \( \epsilon \) is a small constant added to avoid division by zero. Similarly, for each item \( i \), we define:
\begin{equation} \label{eq:pop_direction_item}
    \bar{\mathbf{e}}_{\text{pop}}^{(0)}(i) = \frac{\sum_{u \in \mathcal{N}(i)} b_{ui} \cdot \mathbf{e}_u^{(0)}}{\sum_{u \in \mathcal{N}(i)} b_{ui} + \epsilon}, \quad
    \bar{\mathbf{e}}_{\text{pref}}^{(0)}(i) = \frac{\sum_{u \in \mathcal{N}(i)} (1 - b_{ui}) \cdot \mathbf{e}_u^{(0)}}{\sum_{u \in \mathcal{N}(i)} (1 - b_{ui}) + \epsilon}.
\end{equation}

We define the popularity direction vector of a node as the difference between its popularity and preference centroids:
\begin{equation} \label{eq:pop_direction}
    \mathbf{d}_{\text{pop}}^{(0)}(u) = \bar{\mathbf{e}}_{\text{pop}}^{(0)}(u) - \phi \cdot \bar{\mathbf{e}}_{\text{pref}}^{(0)}(u), \quad
    \mathbf{d}_{\text{pop}}^{(0)}(i) = \bar{\mathbf{e}}_{\text{pop}}^{(0)}(i) - \phi \cdot \bar{\mathbf{e}}_{\text{pref}}^{(0)}(i).
\end{equation}
The popularity direction vector captures how popularity pulls the node embeddings away from preferences. This vector represents the direction from preference to popularity in embedding space. The preference centroid coefficient \(\phi \in [0,1]\) controls the contribution of the preference centroid in the popularity direction and adjusts for potential noise in the preference centroid. In datasets where popularity bias is severe due to heavily skewed data, preference signals may contain unreliable or noisy information. In such cases, using a larger $\phi$ may yield a less accurate estimate of the popularity direction, whereas a smaller $\phi$ can provide a more reliable estimation. In datasets with more diverse interactions and less popularity, preference signals can be more reliable, and a larger $\phi$ enables a more accurate estimate of the popularity direction.

We leverage the vector projection formula~\cite{perwass-springer09} to decompose a vector along the direction of another. By projecting one vector onto another, we can obtain the component of the first vector that lies along the direction of the second~\cite{perwass-springer09}. We project the node embedding onto the popularity direction vector to get the component of an embedding aligned with that direction. By subtracting this component from the node embedding, we remove the popularity component while preserving user preferences, and thus can obtain unbiased embeddings for users and items free from popularity. The final debiased embeddings for user $u$ at layer $0$ can be written as:
\vspace{-0.09cm}
\begin{align} \label{eq:projected_unbiased_user_emb}
\tilde{\mathbf{e}}_u^{(0)} 
    &= \mathbf{e}_u^{(0)} - \text{Proj}_{\mathbf{d}_{\text{pop}}^{(0)}(u)}(\mathbf{e}_u^{(0)}) \nonumber \\
    &= \mathbf{e}_u^{(0)} - \frac{\langle \mathbf{e}_u^{(0)}, \mathbf{d}_{\text{pop}}^{(0)}(u) \rangle}{\| \mathbf{d}_{\text{pop}}^{(0)}(u) \|^2} \cdot \mathbf{d}_{\text{pop}}^{(0)}(u).
\end{align}
Here, $\text{Proj}_{\mathbf{d}_{\text{pop}}^{(0)}(u)}(\mathbf{e}_u^{(0)})$ denotes the projection of the user embedding $\mathbf{e}_u^{(0)}$ onto the popularity direction $\mathbf{d}_{\text{pop}}^{(0)}(u)$, yielding the component of the user’s representation in the direction of popularity. The notation $\langle \cdot, \cdot \rangle$ represents the dot product between two vectors. Similarly, the final debiased embeddings for each item $i$ at layer $0$ can be written as:
\vspace{-0.09cm}
\begin{align} \label{eq:projected_unbiased_item_emb}
\tilde{\mathbf{e}}_i^{(0)} 
    &= \mathbf{e}_i^{(0)} - \text{Proj}_{\mathbf{d}_{\text{pop}}^{(0)}(i)}(\mathbf{e}_i^{(0)}) \nonumber \\
    &= \mathbf{e}_i^{(0)} - \frac{\langle \mathbf{e}_i^{(0)}, \mathbf{d}_{\text{pop}}^{(0)}(i) \rangle}{\| \mathbf{d}_{\text{pop}}^{(0)}(i) \|^2} \cdot \mathbf{d}_{\text{pop}}^{(0)}(i).
\end{align}
\vspace{-0.09cm}

These debiased embeddings are used as input to the propagation layers of the model. The information propagation for layers $l \geq 1$ can be computed using equation~\eqref{eq:a_node_prop}, where \( \mathbf{e}_u^{(0)} := \tilde{\mathbf{e}}_u^{(0)} \) and \( \mathbf{e}_i^{(0)} := \tilde{\mathbf{e}}_i^{(0)} \) are the debiased embeddings at layer $0$. By debiasing the 0th-layer embeddings and propagating them in higher layers, the model propagates unbiased node information to higher layers. Thus, we remove popularity at the 0-th layer to allow unbiased information to flow through higher layers. The full PPD algorithm 
is provided in Subsection~\ref{sec:PPD_alg}
of the Appendix.

Finally, the final embedding of a node is obtained by combining information from all layers using equation~\eqref{eq:a_node_fianl_emb}. The predicted relevance score between user $u$ and item $i$ is the inner product of their final embeddings: $\hat{y}_{ui} = \mathbf{e}_u^\top \mathbf{e}_i$. Based on these scores, the model can recommend the top-$k$ items that user $u$ has not yet interacted with and that are most likely to reflect the user's interests. Our approach is post-hoc, applying bias correction to already learned node representations. Thus, it can correct bias in a deployed model without retraining, updating only the embeddings of a trained model to mitigate popularity bias.

Note that PPD requires only the embeddings from a pre-trained model and can therefore be used as a general post-hoc debiasing strategy for other latent-factor models (e.g., matrix factorization) using the same procedure, except propagating information from unbiased embeddings. We focus on GNN-based CF in our current study because GNNs are state-of-the-art CF models but tend to severely amplify popularity bias through propagation~\cite{zhou-sigir23}, making them an important use case. Exploring post-hoc methods for other latent-factor models can be a good future research direction.
\footnotetext{This work was supervised by Ren Wang and Elena Zheleva.}

%% file: 5-experimental-setup.tex
\section{Experiments}
In this section, we evaluate the performance of our PPD method by formulating the following research questions:
\begin{itemize}[leftmargin=9pt, nosep]
    \item \textbf{RQ1:} How does PPD perform compared with existing popularity bias correction methods?
    \item \textbf{RQ2:} How does PPD improve the performance of popular and niche item groups relative to other methods, and does improving one come at the expense of the other?
    \item \textbf{RQ3:} How does varying the number of GNN layers affect the performance of PPD?
    \item \textbf{RQ4:} How does PPD perform when applied to other GNN-based CF backbones (e.g., SGL~\cite{wu-sigir21})?
    \item  \textbf{RQ5:} How does incorporating personalized preferences into popularity estimation affect the performance of PPD?
    \item \textbf{RQ6:} How does the preference centroid  hyperparameter ($\phi$) affect the performance of PPD?
    \item \textbf{RQ7:} How does the popularity penalty hyperparameter ($\beta$) in personalized preference affect the performance of PPD?
\end{itemize}

\subsection{Experimental setup}
\textbf{Datasets.} We use three real-world datasets: KuaiRec~\cite{gao-cikm22}, Coat~\cite{schnabel-icml16}, and Yahoo! R3~\cite{marlin-recsys09}. We choose these datasets because they provide unbiased test data, which allows us unbiased evaluation of methods. KuaiRec, from Kuaishou short-video logs, provides a fully observed user–item matrix for unbiased evaluation, with nearly all users rating every video, while the training set remains sparse. The test set includes ratings from $1,411$ users on $3,327$ items. Following~\cite{gao-cikm22}, an interaction is considered positive if a user watches a video for more than twice its duration. Coat contains customer ratings from an online coat store, where each user rated $24$ self-selected items in the training set, and the unbiased test set includes ratings from $16$ randomly shown items per user. Yahoo! R3 dataset contains over $300,000$ self-selected user–song ratings by $15,400$ users in the training set. Additionally, a separate unbiased test set is collected by asking $5,400$ users to provide ratings for $10$ randomly presented songs. Ratings are given on a $5$-point scale, and for our experiments, we consider ratings of $3$ or higher as positive feedback. Table~\ref{tab:dataset_stats} presents a detailed overview of the datasets. Although evaluation on other datasets is possible, they generally do not provide unbiased test data. Those datasets construct test sets by sampling from user interactions in the training data~\cite{wei-sigkdd21}, which still follow a popularity-dominated distribution and prevent unbiased evaluation. Therefore, we use the three mentioned datasets, which provide unbiased test data under random or full-exposure settings, varying sizes, sparsity, and popularity, covering diverse real-world scenarios.
\begin{table} [hb]
    \fontsize{8.9}{9}\selectfont
    \centering
    \captionsetup{justification=raggedright, width=1.0\textwidth}
    \caption{Detailed datasets statistics.}
    \label{tab:dataset_stats}
    \vspace{-0.2cm}
    \begin{tabular}{lccc}
        \toprule
        Dataset & KuaiRec & Coat & Yahoo! R3 \\
        \midrule
        \#Users         & 7,176    & 290 & 14,382    \\
        \#Items         & 10,728   & 300 & 1,000     \\
        \#Interactions  & 1,153,106 & 5,490 & 129,748   \\
        Sparsity        &  0.01498  &  0.06303 & 0.00902 \\
        \bottomrule
    \end{tabular}
    \vspace{-0.3em}
\end{table}
\begin{table*} [ht]
    \setlength{\tabcolsep}{2.45pt} 
    \fontsize{8.9}{9}\selectfont
    \centering
    \captionsetup{justification=raggedright, width=1.0\textwidth}
    \caption{A comparison of different methods across datasets and evaluation metrics using LightGCN as the backbone. The best-performing method is shown in bold, and the second-best method is underlined. Performance improvement (\%) is given compared to the best baseline for all evaluation metrics. We additionally report the runtime (in minutes) for each method and the corresponding runtime improvement of PPD in terms of time speedup compared to the best-performing baseline.}
    \label{tab:main_results}
    \vspace{-0.15cm}
    \begin{tabular}{llcccccccccccc}
        \toprule
        Debiasing type & Method & \multicolumn{4}{c}{KuaiRec} & \multicolumn{4}{c}{Coat} & \multicolumn{4}{c}{Yahoo! R3} \\
        \cmidrule(lr){3-6} \cmidrule(lr){7-10} \cmidrule(lr){11-14}
        & & Recall & NDCG & HR & Runtime & Recall & NDCG & HR & Runtime & Recall & NDCG & HR & Runtime \\
        \midrule
        \multirow{2}{*} {Post-hoc} & PPD      & \textbf{0.0988} & \textbf{0.2899} & \textbf{0.9043} & 112.31 & \textbf{0.2407} & \textbf{0.1957} & \textbf{0.5871} & 0.11 & \textbf{0.1505} & \textbf{0.0695} & \textbf{0.2237} & 3.35 \\
        & DAP~\cite{chen-front24}   & 0.0289 & 0.0284 & 0.3007 & 260.13 & 0.1607 & 0.0777 & 0.4555 & 0.28 & 0.1416 & 0.0647 & 0.2150 & 32.25 \\
        \midrule
        None & LightGCN~\cite{he-sigir20} & 0.0055 & 0.0043 & 0.0709 & 115.42 & 0.1572 & 0.0773 & 0.4626 & 2.97 & 0.1403 & 0.0640 & 0.2136 & 13.83 \\
        \midrule
        \multirow{6}{*}{In-training} 
        & APDA~\cite{zhou-sigir23}   & 0.0097 & 0.0077 & 0.1057 & 832.52 & 0.1435 & 0.0727 & 0.4413 & 0.68 & 0.1392 & 0.0627 & 0.2081 & 31.58 \\
        & NAVIP~\cite{kim-cikm22}   & 0.0006 & 0.0039 & 0.0759 & 121.34 & 0.1413 & 0.0730 & 0.4377 & 0.35 & 0.1439 & 0.0648 & 0.2161 & 13.30 \\
        & MACR~\cite{wei-sigkdd21}   & 0.0104 & 0.0107 & 0.1872 & 119.23 & 0.1002 & 0.0451 & 0.3345 & 1.17 & 0.0089 & 0.0033 & 0.0181 & 4.03 \\
        & IPSCN~\cite{joachims-wsdm17}   & 0.0010 & 0.0060 & 0.0574 & 118.32 & \underline{0.2296} & \underline{0.1933} & \underline{0.5765} & 3.15 & 0.1309 & 0.0589 & 0.1977 & 5.12 \\
        & SAM-REG~\cite{boratto-ipm21}   & 0.0018 & 0.0133 & 0.1121 & 132.12 & 0.2206 & 0.1892 & 0.5730 & 3.75 & 0.1226 & 0.0551 & 0.1847 & 7.14 \\
        & PPAC~\cite{ning-www24}   & \underline{0.0345} & \underline{0.0483} & \underline{0.6177} & 260.21 & 0.1444 & 0.0716 & 0.3915 & 0.26 & \underline{0.1462} & \underline{0.0657} & \underline{0.2178} & 11.52 \\
        \midrule
        & Improvement    & +186.4\% & +500.2\% & +46.4\% & 2.32\textit{×} & +4.8\% & +1.2\% & +1.8\% & 28.64\textit{×} & +2.9\% & +5.8\% & +2.7 & 3.44\textit{×} \\
        \bottomrule
    \end{tabular}
\end{table*}

\textbf{Evaluation metrics.}
Following the all-ranking protocol~\cite{krichene-sigkdd20}, we rank all items per user, excluding those with positive feedback in the training set. Recommendation performance is measured using Recall@$k$, NDCG@$k$, and Hit Ratio (HR@$k$) with $k=20$ by default, where $k$ is the evaluation cutoff. The KuaiRec dataset presents a special case. Its test set is constructed from a dense subset of the user-item matrix~\cite{gao-cikm22}, rather than being a random sample from the entire space. As a result, the traditional all-ranking protocol is not applicable. Instead, for KuaiRec dataset, we evaluate by ranking only the $3,327$ items that are fully exposed.

\textbf{Baselines.} We evaluate the performance of PPD relative to popularity bias correction methods in CF, including post-hoc method DAP~\cite{chen-front24}, interaction-weighting methods applied during aggregation like APDA~\cite{zhou-sigir23} and NAVIP~\cite{kim-cikm22}, causal inference-based approaches such as MACR~\cite{wei-sigkdd21} and PPAC~\cite{ning-www24}, IPW  methods like IPSCN~\cite{gruson-wsdm19}, and a regularization-based approach SAM-REG~\cite{boratto-ipm21}.
We include LightGCN~\cite{he-sigir20} as a standard GNN-based CF backbone without any debiasing, using it as the backbone for all baselines and for our method. These baselines cover all the major classes of debiasing strategies, while detailed descriptions of baselines are provided in the Appendix~\ref{sec:baselines}. Since PPD focuses on pure CF using only interaction data, without any user or item features or side information, we include baselines that are adaptable to this setting. We provide the detailed implementation details in the Appendix~\ref{sec:imp_details}. Our GitHub code is publicly available in the provided link\footnote{Code is available at \url{https://github.com/edgeslab/Posthoc_Popularity_Debiasing}}.

%% file: 6-results.tex
\subsection{Results}
In this section, we present the results of our method with respect to the research questions introduced earlier.

\begin{table*} [hb]
    \setlength{\tabcolsep}{2.9pt} 
    \fontsize{8.9}{9}\selectfont
    \centering
    \captionsetup{justification=raggedright, width=1.0\textwidth}
    \caption{A comparison of different methods across datasets using LightGCN as the backbone, evaluated on the top $\bm{20\%}$ and bottom $\bm{80\%}$ of items sorted by interaction frequency in the training data. The best-performing method is shown in bold, and the second-best method is underlined. Performance improvement (\%) is given compared to the best baseline.}
    \label{tab:recall_ndcg_top_bottom}
    \vspace{-0.15cm}
    \begin{tabular}{llcccccccccccc}
        \toprule
        \multirow{2}{*}{Debiasing type} & \multirow{2}{*}{Method} 
        & \multicolumn{4}{c}{KuaiRec} & \multicolumn{4}{c}{Coat} & \multicolumn{4}{c}{Yahoo! R3} \\
        \cmidrule(lr){3-6} \cmidrule(lr){7-10} \cmidrule(lr){11-14}
        & & \multicolumn{2}{c}{Top 20\%} & \multicolumn{2}{c}{Bottom 80\%} 
          & \multicolumn{2}{c}{Top 20\%} & \multicolumn{2}{c}{Bottom 80\%} 
          & \multicolumn{2}{c}{Top 20\%} & \multicolumn{2}{c}{Bottom 80\%} \\
        & & Recall & NDCG & Recall & NDCG & Recall & NDCG & Recall & NDCG & Recall & NDCG & Recall & NDCG \\
        \midrule
        \multirow{2}{*}{Post-hoc} 
        & PPD & \textbf{0.1167} & \textbf{0.2931} & \textbf{0.0010} & \textbf{0.0005} 
              & \textbf{0.3770} & \textbf{0.2259} & \textbf{0.1550} & \textbf{0.1291} 
              & \textbf{0.3222} & \textbf{0.1450} & \textbf{0.0352} & \underline{0.0127} \\
        & DAP~\cite{chen-front24} & 0.0337 & 0.0297 & 0.0000 & 0.0000 
              & 0.3627 & 0.1524 & 0.1215 & 0.0582 
              & 0.2890 & 0.1267 & 0.0337 & 0.0123 \\
        \midrule
        None & LightGCN~\cite{he-sigir20} 
              & 0.0064 & 0.0027 & 0.0000 & 0.0000 
              & 0.3671 & 0.1565 & 0.0643 & 0.0237 
              & 0.2885 & 0.1263 & 0.0332 & 0.0118 \\
        \midrule
        \multirow{6}{*}{In-training} 
        & APDA~\cite{zhou-sigir23}   
              & 0.0116 & 0.0085 & 0.0000 & 0.0000 
              & 0.3073 & 0.1299 & 0.0714 & 0.0286 
              & 0.2643 & 0.1147 & \underline{0.0344} & \textbf{0.0129} \\
        & NAVIP~\cite{kim-cikm22}   
              & 0.0003 & 0.0003 & 0.0000 & 0.0000 
              & 0.3421 & 0.1395 & 0.0462 & 0.0201 
              & \underline{0.3008} & \underline{0.1295} & 0.0327 & 0.0125 \\
        & MACR~\cite{wei-sigkdd21}  
              & 0.0125 & 0.0111 & 0.0000 & 0.0000 
              & 0.1724 & 0.0539 & 0.0663 & 0.0320 
              & 0.0003 & 0.0001 & 0.0142 & 0.0049 \\
        & IPSCN~\cite{joachims-wsdm17} 
              & 0.0003 & 0.0012 & \underline{0.0003} & \underline{0.0001} 
              & \underline{0.3672} & \underline{0.2238} & 0.1509 & \underline{0.1277} 
              & 0.2697 & 0.1177 & 0.0297 & 0.0116 \\
        & SAM-REG~\cite{boratto-ipm21} 
              & 0.0023 & 0.0132 & 0.0001 & \underline{0.0001} 
              & 0.3341 & 0.2139 & \underline{0.1531} & 0.1272 
              & 0.2775 & 0.1191 & 0.0116 & 0.0044 \\
        & PPAC~\cite{ning-www24}    
              & \underline{0.0407} & \underline{0.0494} & 0.0000 & 0.0000 
              & 0.3627 & 0.1595 & 0.0731 & 0.0327 
              & 0.1462 & 0.0657 & 0.0336 & 0.0124 \\
        \midrule
        & Improvement & +186.7\% & +492.3\% & +233.3\% & +400\%
        & +2.7\% & +0.94\% & +1.24\% & +1.09\%
        & +7.1\% & +12.0\% & +2.33\% & -1.5\% \\

        \bottomrule
    \end{tabular}
    \vspace{-0.21cm}
\end{table*}
\textbf{Performance comparison between our method and the baselines (RQ1).}
In response to RQ1, we evaluate whether PPD outperforms existing popularity bias correction approaches. Table~\ref{tab:main_results} compares PPD with baselines, where in-training methods correct bias during training and post-hoc methods do so after training. PPD consistently achieves the best results across all metrics. On KuaiRec, where popularity bias can be severe (Figure~\ref{fig:pop_bias_demonstration}), PPD outperforms the strongest baseline (PPAC) with relative improvements of 186.4\% in Recall@20, 500.2\% in NDCG@20, and 46.4\% in HR@20, highlighting its effectiveness in highly skewed data. On Coat, with milder popularity (Figure~\ref{fig:pop_bias_demonstration}), PPD outperforms the strongest baseline IPSCN by 4.8\% in Recall@20, 1.2\% in NDCG@20, and 1.8\% in HR@20, remaining competitive in less biased scenarios. On sparse Yahoo! R3 (Table~\ref{tab:dataset_stats}), PPD exceeds all baselines, improving 2.9\%, 5.8\%, and 2.7\% over PPAC. PPD also outperforms the post-hoc baseline DAP~\cite{chen-front24}, as PPD  estimates interaction-level popularity for debiasing by considering both global and personalized preferences, rather than clustering similar nodes and relying on simple node degree information. Moreover, PPD consistently outperforms aggregation-weighting methods such as APDA~\cite{zhou-sigir23} and NAVIP~\cite{kim-cikm22}, supporting our claim that applying weights in every training epoch can distort model learning and does not effectively reduce popularity bias.
These results demonstrate several key insights. First, the substantial improvements on KuaiRec emphasize that PPD is particularly effective in settings where popularity can be most prominent, confirming its robustness to extreme skew in interaction frequency. Second, the consistent improvements on Coat indicate that PPD also generalizes well to datasets with less pronounced popularity bias. Finally, even on Yahoo! R3, where extremely sparse interactions can make debiasing particularly difficult due to fewer preference signals in the data, PPD yields consistent improvements, suggesting its ability to correct for bias in sparser datasets. Together, these findings show that PPD consistently outperforms the baselines across datasets with varying levels of skewness and sparsity.

In addition to performance, PPD demonstrates computational advantages over existing methods because it does not require retraining the model. As shown in Table~\ref{tab:main_results}, PPD is consistently faster than all methods across all datasets. It achieves 2.32\textit{×}, 28.64\textit{×}, and 3.44\textit{×} speedups on KuaiRec, Coat, and Yahoo! R3, respectively, compared to the best-performing baselines. These results demonstrate that PPD not only improves recommendation performance by reducing popularity bias but also lowers computational cost. Moreover, PPD requires only the embeddings from the pre-trained model, avoiding the need to load and process the training data or train the GNN model on the GPU. As a result, it requires less GPU memory. PPD requires approximately 401MB, 228MB, and 352MB of GPU memory on KuaiRec, Coat, and Yahoo! R3, respectively, which is approximately half of that used by the best-performing baselines.
\begin{figure*} [ht]
  \centering
  \captionsetup{justification=raggedright, margin=0cm}
  \captionsetup[subfigure]{labelfont=normalfont, textfont=normalfont}

  \begin{subfigure}{0.28\textwidth}
    \includegraphics[width=\linewidth]{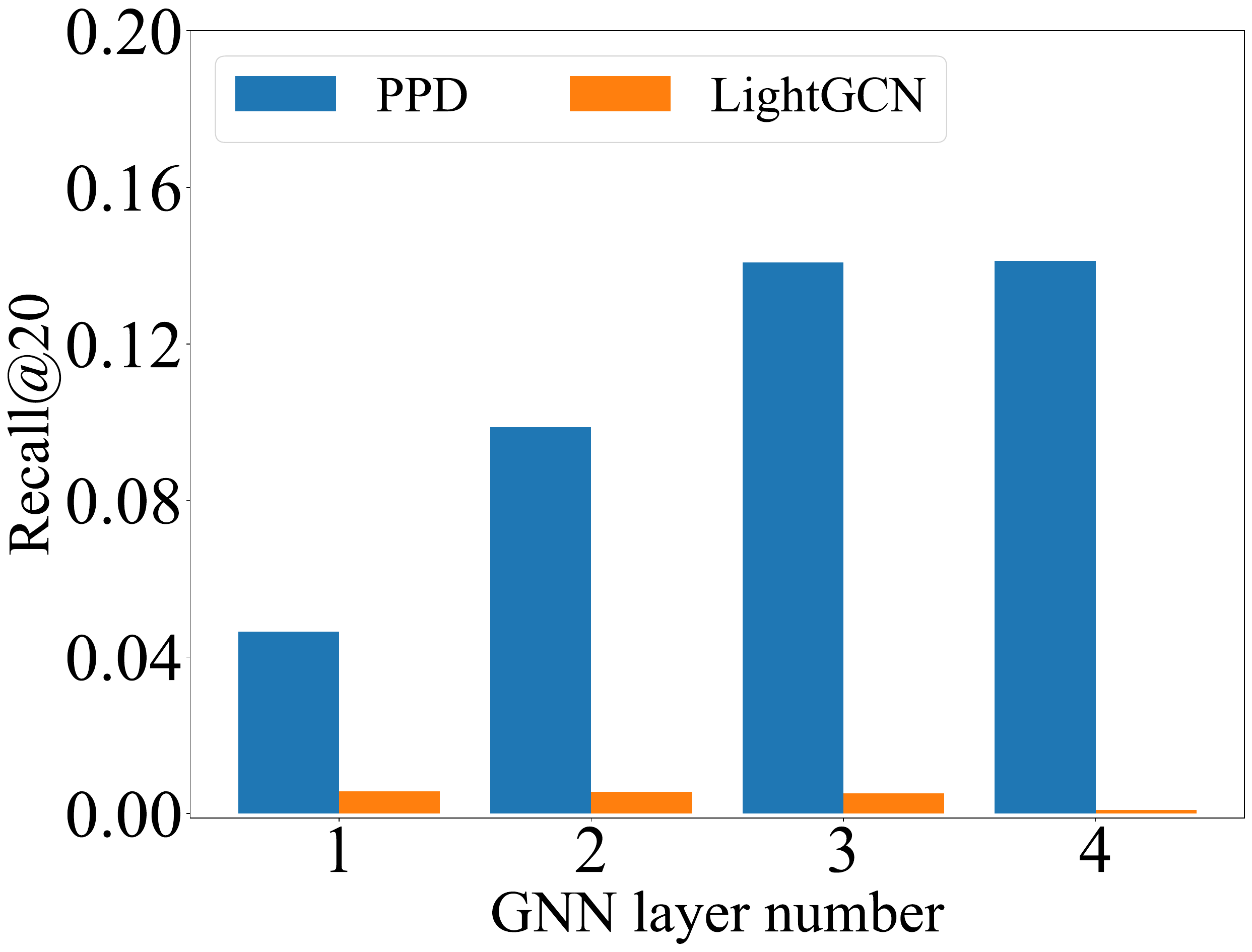}
    \caption{KuaiRec dataset} \label{fig:kuairec_layer_variation_recall}
  \end{subfigure}
  \begin{subfigure}{0.28\textwidth}
    \includegraphics[width=\linewidth]{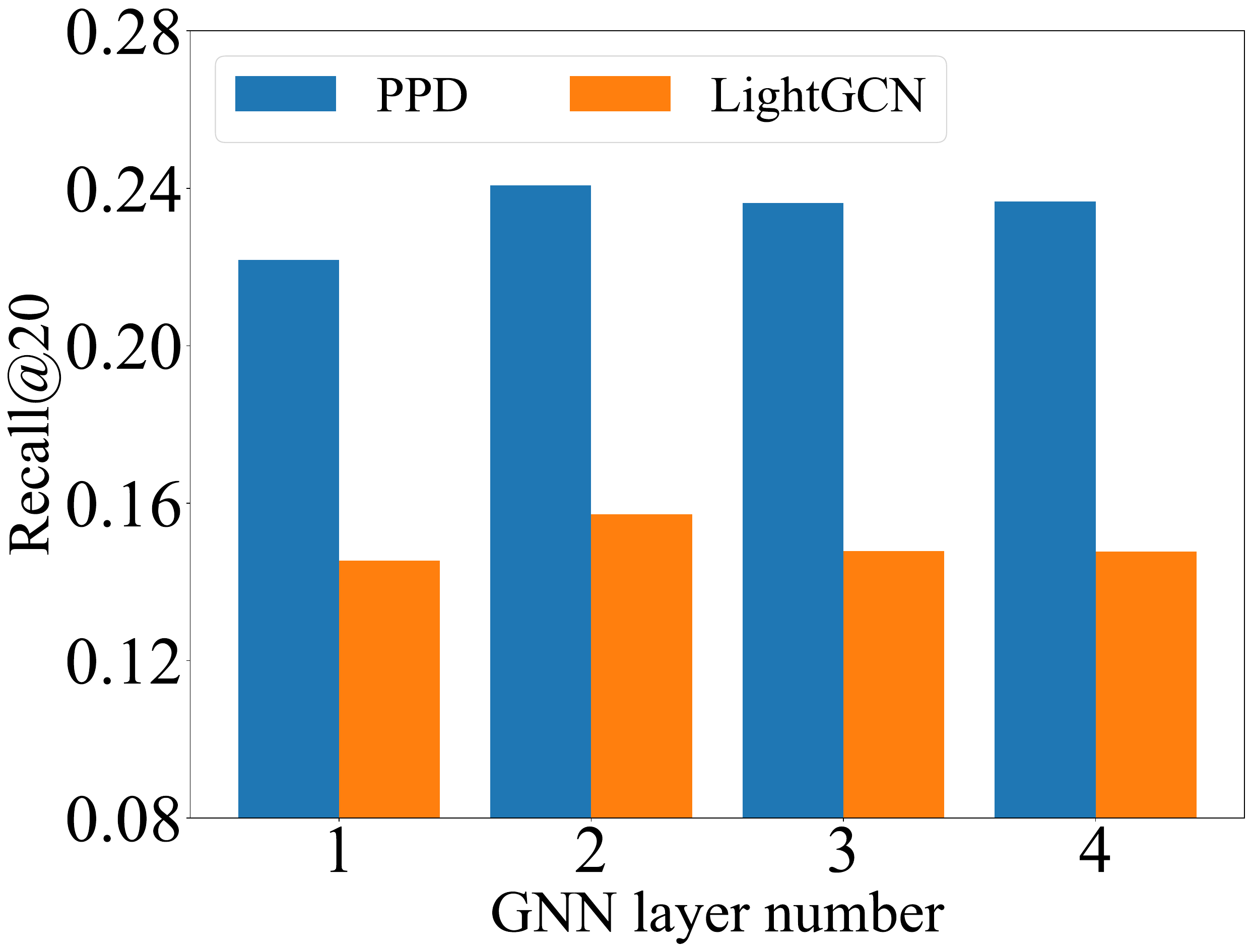}
    \caption{Coat dataset} \label{fig:coat_layer_variation_recall}
  \end{subfigure}
  \begin{subfigure}{0.28\textwidth}
    \includegraphics[width=\linewidth]{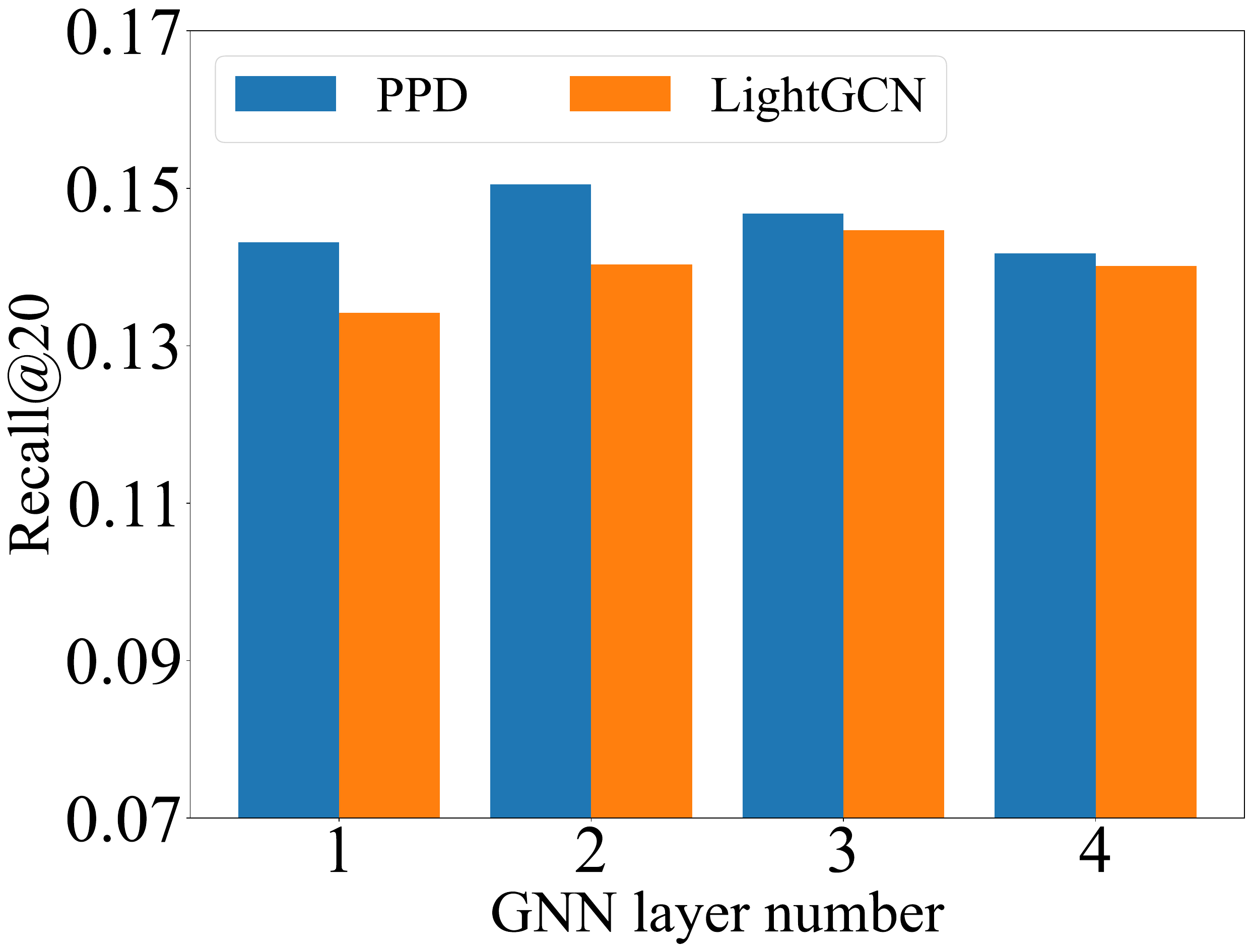}
    \caption{Yahoo! R3 dataset} \label{fig:yahoo_layer_variation_recall}
  \end{subfigure}
  \vspace{-0.2cm}

  \captionsetup{width=0.98\textwidth}
  \caption{Effect of varying the number of layers on performance (Recall@$\bm{20}$) for PPD and LightGCN.\label{fig:layer_variations_recall}}
  \vspace{-0.2cm}
\end{figure*}

\textbf{Popular and niche item performance improvement (RQ2).}
RQ2 evaluates performance on popular (head) and niche (tail) items, examining whether improvements on one group come at the expense of the other. Popularity bias correction can create such a trade-off, where enhancing tail-item performance may degrade head-item accuracy~\cite{zhou-sigir23}. To investigate this, following~\cite{zhou-sigir23, chen-front24}, we divide items into the top 20\% most frequently interacted ones (popular) and the remaining 80\% (niche) based on the training data, and evaluate performance on these subsets. This allows assessment of whether PPD balances recommendation quality for both head and tail items. We choose the 20\%-80\% split, as it is commonly used in recommender systems to analyze popularity~\cite{zhou-sigir23, chen-front24}, reflecting power-law popularity distributions. As shown in Table~\ref{tab:recall_ndcg_top_bottom}, PPD consistently achieves the best performance across both groups. On KuaiRec, PPD improves head performance by +186.7\% in Recall@20 and +492.3\% in NDCG@20 compared to the best baseline. For niche items, where all methods struggle due to very skewed data in tails, PPD still shows the best performance, achieving relative improvements of +233.3\% in Recall@20 and +400\% in NDCG@20 compared to best baseline. The overall low scores on KuaiRec tails stem from the extremely imbalanced distribution, where most niche items may have only a few interactions, making them inherently difficult to recommend. On Coat, PPD also improves both head and niche groups simultaneously. For head items, it yields +2.7\% Recall@20 and +0.94\% NDCG@20 over the best baseline. For niche items, it further improves by +1.24\% Recall@20 and +1.09\% NDCG@20, showing that gains on tail recommendations do not compromise head performance. On Yahoo! R3, PPD improves head recommendations with +7.1\% Recall@20 and +12.0\% NDCG@20 compared to best baseline, while also improving tail performance with +2.33\% Recall@20, though NDCG shows a slight drop (–1.5\%) compared to best baseline. Across all datasets, PPD achieves the highest HR@20 compared to the baselines, though we do not report the results due to space constraints.
In summary, PPD enhances recommendations for tail items, demonstrating its ability to promote niche items while maintaining strong performance for head items. PPD explicitly removes popularity-driven components from embeddings while retaining preference-driven signals through interaction-level popularity scores. This ensures that head items are recommended when they genuinely reflect user interests, and tail items are recommended when aligned with user preferences. Therefore, the typical head–tail trade-off is avoided: head accuracy is preserved, and tail performance is improved.

\textbf{Effect of GNN layer depth on model performance (RQ3).}
We investigate how varying the number of GNN layers influences the performance of the PPD method. Figure~\ref{fig:layer_variations_recall} demonstrates the effect of increasing the number of layers. On the KuaiRec dataset, where popularity bias is most severe, LightGCN reaches peak performance with two layers but suffers degradation as more layers are added, indicating that deeper propagation amplifies popularity bias and over-smoothing. In contrast, PPD continues to benefit from additional layers, suggesting that its debiasing mechanism can mitigate bias amplification while leveraging multi-hop information. On the Coat dataset, where popularity bias is less pronounced, both models achieve their best performance with two layers, and deeper propagation beyond this point may introduce noise rather than useful signals, reducing accuracy. For Yahoo! R3, PPD achieves optimal results with two layers, while LightGCN slightly peaks at three layers, reflecting that in moderately biased and sparse settings, LightGCN can still gain from slightly deeper propagation. Importantly, across all datasets and all depths, PPD consistently outperforms LightGCN, demonstrating robustness and resilience to over-smoothing and bias mitigation. 
For NDCG@20, PPD also outperforms across different layer depths, exhibiting trends similar to Recall@20.
These findings highlight that the optimal number of layers depends on the dataset, and PPD maintains stronger performance by mitigating popularity bias across different depths. 

\textbf{RQ4-RQ7.} Beyond all these main results from RQ1-RQ3, we also evaluate four additional analysis in RQ4-RQ7. We present detailed results and analysis for RQ4–RQ7 in Appendix~\ref{sec:add_exp}.
In RQ4, we apply PPD to another stronger GNN-based CF backbone, Self-supervised Graph Learning (SGL), to evaluate the generalizability of our debiasing strategy. PPD consistently outperforms SGL across all three datasets (see Table~\ref{tab:ppd_sgl_results}), with particularly large gains on the popularity-dominated KuaiRec dataset, demonstrating its effectiveness on different GNN-based CF backbones.
In RQ5, we examine the role of personalized preference ($r_{ui}$) when estimating the popularity score ($b_{ui}$). The comparison between PPD and its variant without personalized preference (PPD w/o-PP) shows that removing this component when estimating the popularity score reduces performance across datasets (see Table~\ref{tab:ppd_without_rui}), as relying solely on global preference risks discarding genuinely relevant popular items. This confirms that personalized preference is crucial for isolating the residual effect of popularity.
In RQ6, we study the effect of the preference centroid coefficient ($\phi$) hyperparameter, which balances the contribution of preference signals in popularity direction. The results reveal that the optimal $\phi$ is dataset-dependent. Moderate values (e.g., $\phi = 0.5$) work best when popularity is high and preference signals are weak, as in KuaiRec. Higher values yield better performance in less popularity-skewed datasets where preferences are more reliable, as in Coat and Yahoo! R3 (see Figure~\ref{fig:phi_variations_recall}). These findings highlight the importance of tuning the preference centroid and adapting $\phi$ to each dataset for popularity debiasing. In RQ7, we examine the effect of the popularity penalty coefficient ($\beta$) on PPD during personalized preference estimation. On KuaiRec, Recall@20 improves as $\beta$ increases up to 0.2 but declines at 0.3, while Coat achieves its best performance at $\beta = 0.1$ and Yahoo! R3 at $\beta = 0.3$ (see Figure~\ref{fig:beta_variations_recall}). The results also show that without a penalty ($\beta = 0.0$), the model fails to reach optimal performance, highlighting the need for a popularity penalty in estimating personalized preferences.

%% file: 7-conclusion.tex
\section{Conclusion}
In this work, we propose a popularity debiasing method in a post-hoc manner, a novel method to mitigate popularity bias in GNN-based CF by estimating interaction-level popularity and leveraging it to derive a popularity direction to remove popularity-driven components from node embeddings. Unlike existing approaches, our method requires no retraining and can be applied directly to pre-trained models. Extensive unbiased evaluations on three real-world datasets show that our method consistently outperforms state-of-the-art methods, improving both head and tail recommendations. We further demonstrate that our method generalizes to other GNN-based CF backbones. Future research can extend post-hoc methods to other types of bias, such as exposure bias or position bias, and explore applications beyond GNN-based CF, including other latent-factor models, as well as content-based and dynamic recommendation settings.

%% file: 8-appendix.tex
\section{More Details of Our Method} \label{sec:method_details}
\subsection{PPD algorithm} \label{sec:PPD_alg}
The complete procedure of our proposed PPD method is outlined in Algorithm~\ref{alg:ppd}. The algorithm first estimates interaction-level popularity scores, then constructs a popularity direction vector, and finally removes the effect of popularity via projection.
\begin{algorithm}
\caption{Post-hoc Popularity Debiasing (PPD)}
\label{alg:ppd}
\begin{algorithmic}[1]
\State \textbf{Input:} User embeddings $\{\mathbf{e}^{(0)}_u\}$, item embeddings $\{\mathbf{e}^{(0)}_i\}$, interaction set $\mathcal{E}$, popularity penalty parameter $\beta$, preference centroid coefficient $\phi$, small constant $\epsilon$
\State \textbf{Output:} Debiased embeddings $\{\tilde{\mathbf{e}}^{(0)}_u\}, \{\tilde{\mathbf{e}}^{(0)}_i\}$

\State
\State \textbf{Step 1: Popularity score estimation}
\For{each interaction $(u,i) \in \mathcal{E}$}
    \State Compute global preference score $p_i$ using Eq.~\eqref{eq:global_popularity_rule}
    \State Compute personalized preference score $r_{ui}$ using Eq.~\eqref{eq:personalized_pref_rule}
    \State Compute popularity bias score $b_{ui}$ using Eq.~~\eqref{eq:pop_bias_scores}
\EndFor

\State
\State \textbf{Step 2: Popularity direction construction}
\For{each node $v \in \mathcal{U} \cup \mathcal{I}$}
    \State Compute popularity centroid $\bar{\mathbf{e}}_{\text{pop}}^{(0)}(v)$ and preference centroid $\bar{\mathbf{e}}_{\text{pop}}^{(0)}(v)$ using Eq.~\eqref{eq:pop_direction_user}--\eqref{eq:pop_direction_item}
    \State Compute popularity direction vector $\mathbf{d}_{\text{pop}}^{(0)}(v)$ using Eq.~\eqref{eq:pop_direction}
\EndFor

\State
\State \textbf{Step 3: Debiasing via projection}
\For{each node $v \in \mathcal{U} \cup \mathcal{I}$}
    \State Update node embeddings $\tilde{\mathbf{e}}_v^{(0)} $ by removing popularity-aligned component using Eq.~\eqref{eq:projected_unbiased_user_emb}--\eqref{eq:projected_unbiased_item_emb}
\EndFor

\State \textbf{return} Debiased embeddings $\{\tilde{\mathbf{e}}^{(0)}_u\}, \{\tilde{\mathbf{e}}^{(0)}_i\}$
\end{algorithmic}
\end{algorithm}

\begin{table*} [htpb]
    \fontsize{8.9}{9}\selectfont
    \centering
    \captionsetup{justification=raggedright, width=1.0\textwidth}
    \caption{A comparison of PPD and SGL using SGL as the backbone across datasets and evaluation metrics. The best-performing method is shown in bold. Performance improvement (\%) is given compared to SGL.}
    \label{tab:ppd_sgl_results}
    \vspace{-0.15cm}
    \begin{tabular}{lccccccccc}
        \toprule
        Method & \multicolumn{3}{c}{KuaiRec} & \multicolumn{3}{c}{Coat} & \multicolumn{3}{c}{Yahoo! R3} \\
        \cmidrule(lr){2-4} \cmidrule(lr){5-7} \cmidrule(lr){8-10}
        & Recall & NDCG & HR & Recall & NDCG & HR & Recall & NDCG & HR \\
        \midrule
        PPD & \textbf{0.1352} & \textbf{0.3612} & \textbf{0.9978} & \textbf{0.2437} & \textbf{0.2176} & \textbf{0.6121} & \textbf{0.1381} & \textbf{0.0601} & \textbf{0.2031} \\
        SGL~\cite{wu-sigir21} & 0.0121 & 0.0189 & 0.2574 & 0.2304 & 0.2028 & 0.5801 & 0.1277 & 0.0583 & 0.1909 \\
        \midrule
        Improvement & +1017.4\% & +1811.1\% & +287.6\% & +5.8\% & +7.3\% & +5.6\% & +8.1\% & +3.1\% & +6.4\% \\
        \bottomrule
    \end{tabular}
\end{table*}
\begin{table*}[hb]
    \fontsize{8.9}{9}\selectfont
    \centering
    \captionsetup{justification=raggedright, width=1.0\textwidth}
    \caption{A comparison of PPD and PPD without personalized preference ($r_{ui}$) when estimating the popularity score ($b_{ui}$) using LightGCN as the backbone. The best-performing variant is shown in bold.}
    \label{tab:ppd_without_rui}
    \vspace{-0.15cm}
    \begin{tabular}{lccccccccc}
        \toprule
        Method & \multicolumn{3}{c}{KuaiRec} & \multicolumn{3}{c}{Coat} & \multicolumn{3}{c}{Yahoo! R3} \\
        \cmidrule(lr){2-4} \cmidrule(lr){5-7} \cmidrule(lr){8-10}
        & Recall & NDCG & HR & Recall & NDCG & HR & Recall & NDCG & HR \\
        \midrule
        PPD & \textbf{0.0988} & \textbf{0.2899} & \textbf{0.9043} & \textbf{0.2407} & \textbf{0.1957} & \textbf{0.5871} & \textbf{0.1505} & \textbf{0.0695} & \textbf{0.2237} \\
        PPD w/o-PP & 0.0303 & 0.1348 & 0.7511 & 0.2287 & 0.1921 & 0.5694 & 0.1500 & 0.0689 & 0.2221 \\
        \bottomrule
    \end{tabular}
\end{table*}

\section{Experimental Details} \label{sec:exp_details}
\subsection{Baselines} \label{sec:baselines}
We evaluate our method against popularity bias correction baselines in CF. We provide a description for each of the baselines.
\begin{itemize}[leftmargin=9pt, nosep]
    \item $\textbf{DAP:}$ The Debias the Amplification of Popularity~\cite{chen-front24} method clusters similar nodes, estimates their amplification effect, and removes it from node embeddings to reduce popularity bias.
    \item $\textbf{APDA:}$ The Adaptive Popularity Debiasing Aggregator~\cite{zhou-sigir23} is a popularity bias correction method in GNN-based CF that adaptively apply per-edge weights using an inverse popularity score to mitigate popularity bias during aggregation.
    \item $\textbf{NAVIP:}$ The Neighborhood Aggregation via Inverse Propensity (NAVIP)~\cite{kim-cikm22} method is a GNN-based approach that mitigates popularity bias by using inverse propensity scores as edge weights to debias neighbor aggregation.
    \item $\textbf{MACR:}$ The Model-Agnostic Counterfactual Reasoning~\cite{wei-sigkdd21} uses multi-task learning and counterfactual inference to isolate and remove the direct effect popularity.
    \item $\textbf{IPSCN:}$ Based on the principle of the generic IPW method~\cite{joachims-wsdm17}, this method~\cite{gruson-wsdm19} also introduces max-capping and normalization of the propensity values to reduce their variance.
    \item $\textbf{SAM-REG:}$ The Sampling and Regularization~\cite{boratto-ipm21} method balances samples across popular and less popular items and regularizes against relevance–popularity correlation to mitigate popularity bias.
    \item $\textbf{PPAC:}$ The Personal Popularity Aware Counterfactual~\cite{ning-www24} framework integrates both personal and global popularity using counterfactual inference to mitigate popularity bias.
    \item $\textbf{LightGCN:}$ We train a LightGCN~\cite{he-sigir20} model without applying any bias correction. This model serves as a backbone for GNN-based CF and does not account for popularity bias.
\end{itemize}

\subsection{Implementation details} \label{sec:imp_details}
Since LightGCN~\cite{he-sigir20} is the most widely used backbone for GNN-based CF, we adopt it for all baselines and our method, unless otherwise specified. The model is implemented in PyTorch and optimized using Adam. We apply early stopping with a patience of $100$ epochs, stopping if Recall@$20$ on the validation set does not improve for $100$ consecutive epochs. This setting allows the model to achieve its best performance while maintaining the early stopping criterion. We tune the learning rate among $1{e}^{-3}, 1{e}^{-4}, 1{e}^{-5}$ and the regularization coefficient among $0.0, 1{e}^{-3}, 1{e}^{-5}$. Batch size is $2{,}048$ and embedding dimension is $256$. All other parameters follow the default settings of LightGCN. Unless stated otherwise, we use a two-layer GCN that integrates up to two-hop neighbors. Popularity scores ($b_{ui}$) are computed by min–max normalizing global preference ($p_i$) and personalized preference ($r_{ui}$). Hyperparameters for our method are tuned with popularity penalty coefficient $\beta \in \{0.0, 0.1, 0.2, 0.3\}$ and preference centroid coefficient $\phi \in \{0.0, 0.25, 0.50, 0.75, 1.0\}$. $\beta$ is restricted up to 0.3 to avoid over-penalizing popularity, which could suppress preference signals and weaken the estimation of personalized preference. The small constant $\epsilon$ in equations~\eqref{eq:pop_direction_user} and~\eqref{eq:pop_direction_item} is set to $1{e}^{-8}$. For unbiased validation, we split the unbiased test data of each original dataset, using one-third for validation and the remaining two-thirds for testing for KuaiRec and Coat. However, for Yahoo! R3, the unbiased test set contains a very small number of interactions on average. Therefore, we do not split it and directly use it as the test set. The validation dataset for Yahoo! R3 is constructed by sampling interactions from the training set. We use the dot product as the ${sim}(\cdot)$ function in equations~\eqref{eq:global_popularity_rule} and~\eqref{eq:personalized_pref_rule}. To validate our method on other GNN-based CF models, we also use SGL~\cite{wu-sigir21} as a backbone (see RQ4). 
Following~\cite{zhang-neurips23}, in our experiments, we set the contrastive loss temperature to $0.15$ and its weight to $0.2$ for SGL, while keeping all other settings at the default LightGCN settings. 
\begin{figure*}
  \centering
  \captionsetup{justification=raggedright, margin=0cm}
  \captionsetup[subfigure]{labelfont=normalfont, textfont=normalfont}

  \begin{subfigure}{0.28\textwidth}
    \includegraphics[width=\linewidth]{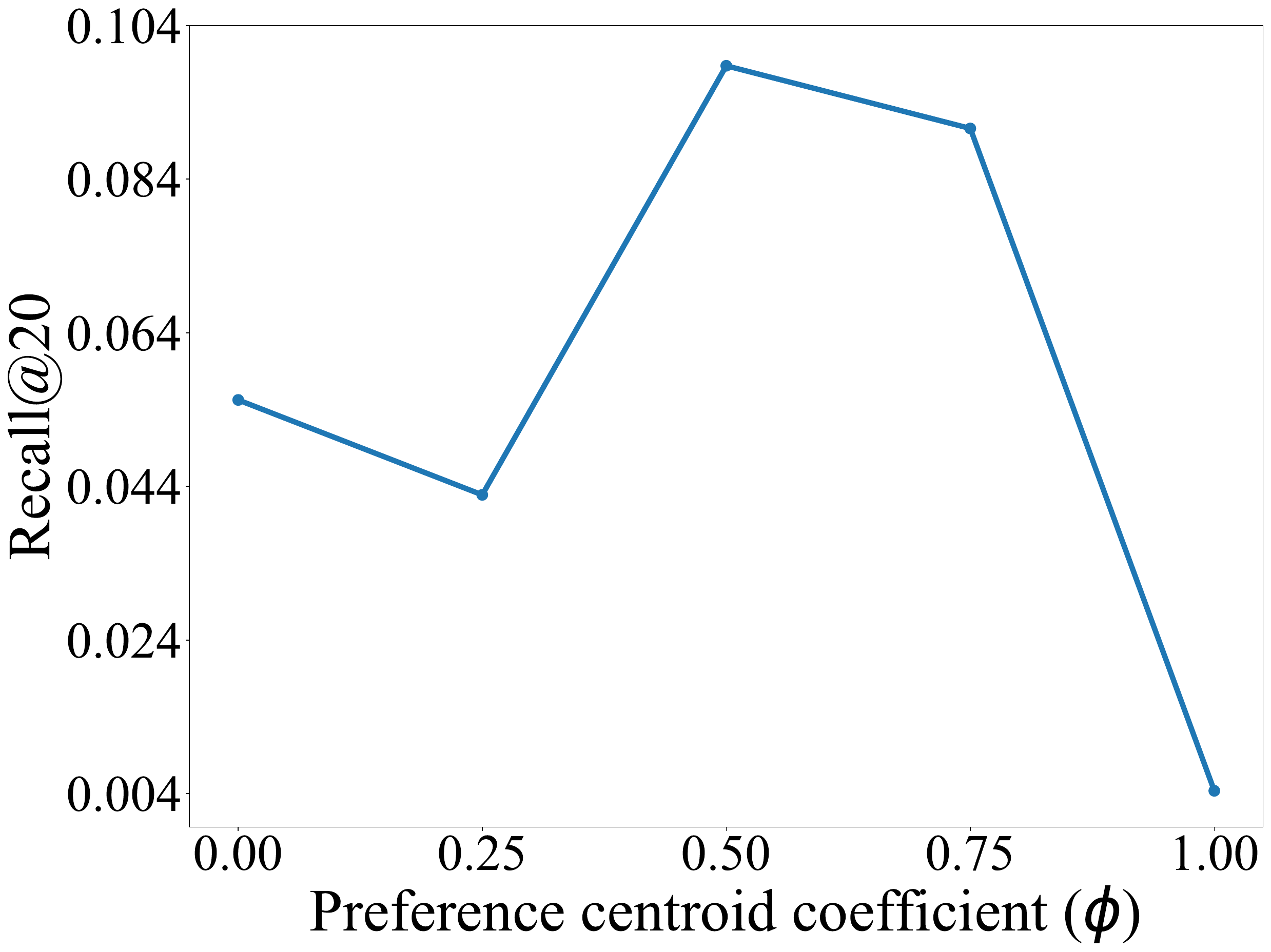}
    \caption{KuaiRec dataset}
  \end{subfigure}
  \begin{subfigure}{0.28\textwidth}
    \includegraphics[width=\linewidth]{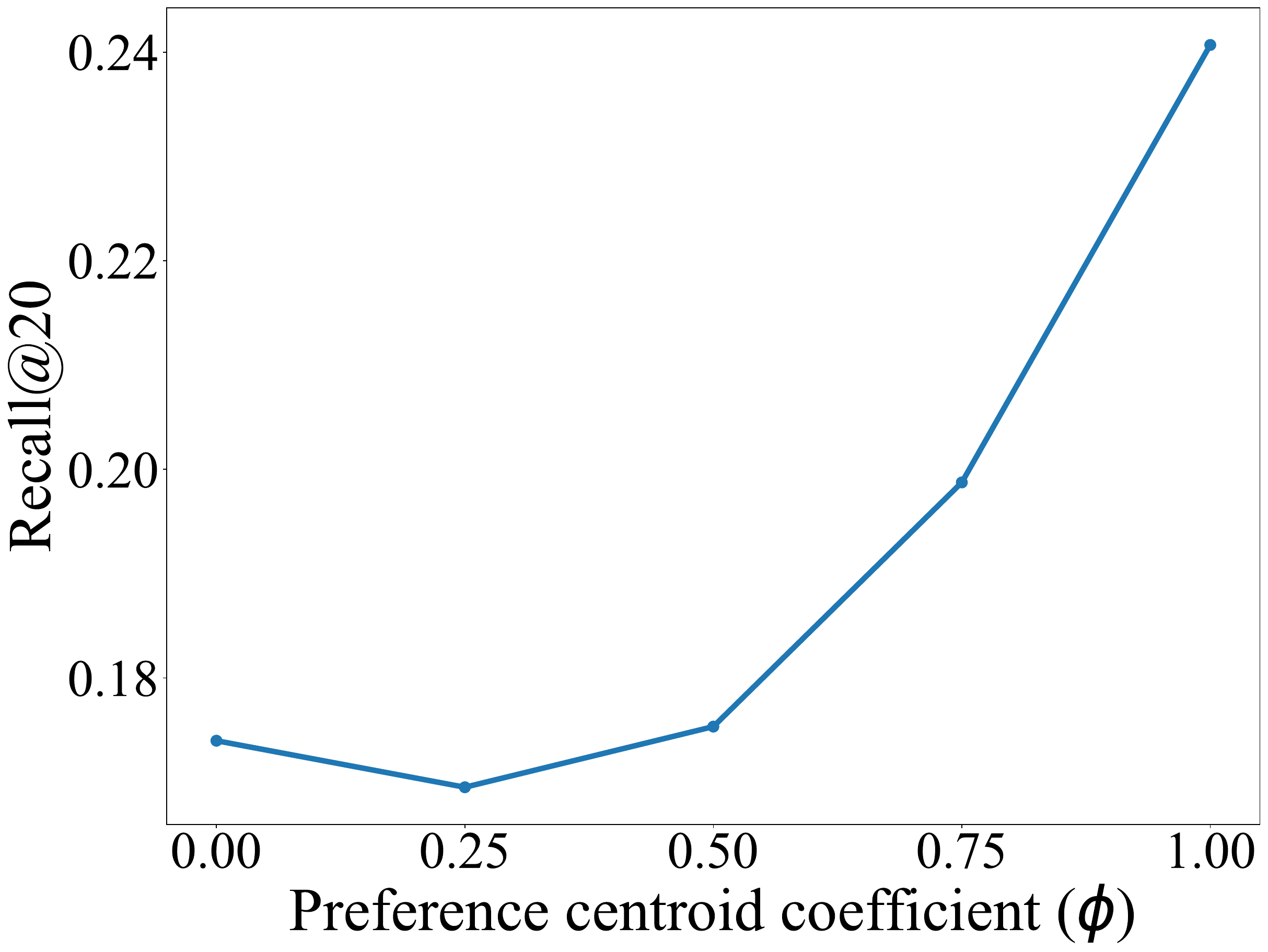}
    \caption{Coat dataset}
  \end{subfigure}
  \begin{subfigure}{0.28\textwidth}
    \includegraphics[width=\linewidth]{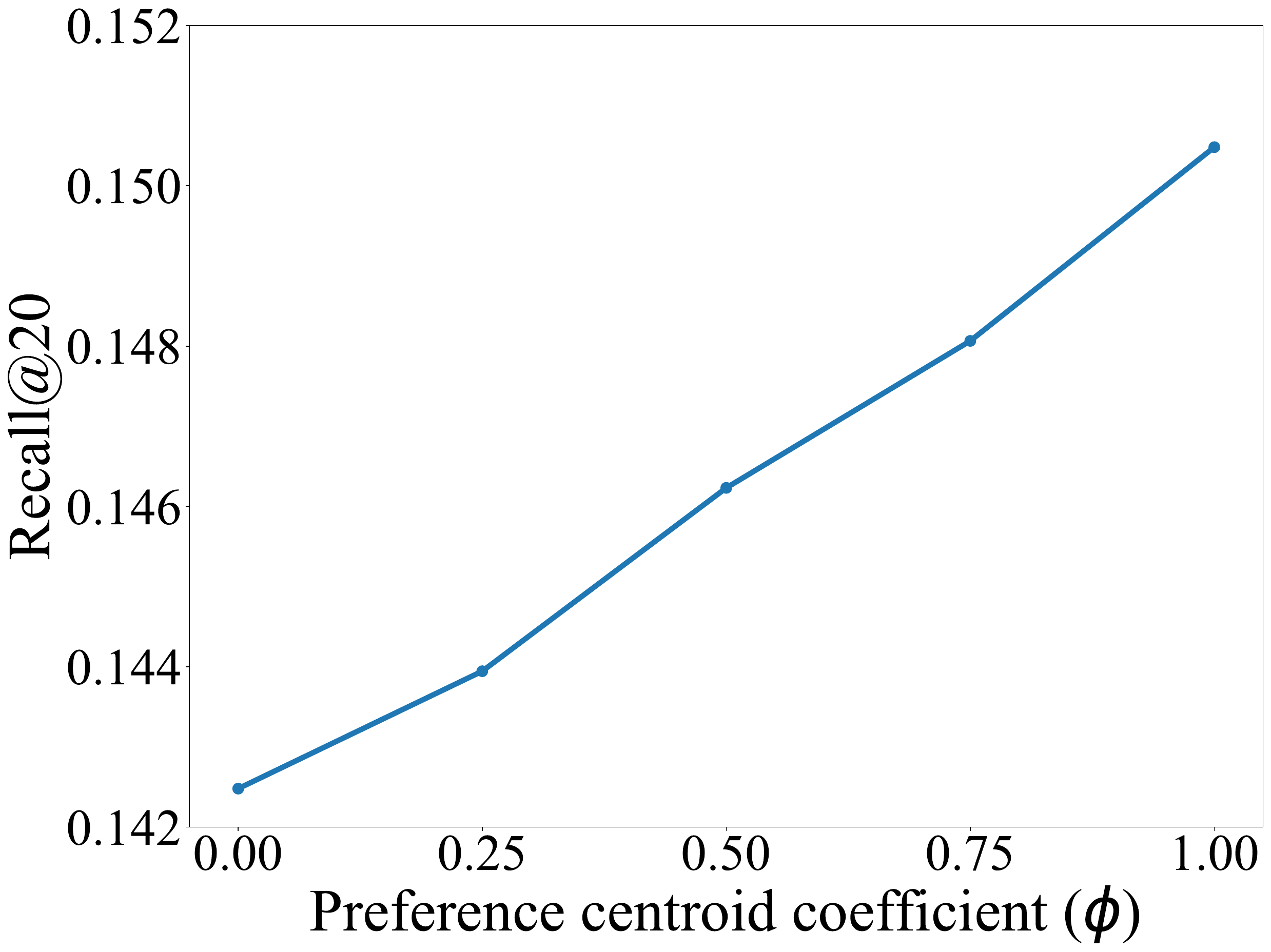}
    \caption{Yahoo! R3 dataset}
  \end{subfigure}
  \vspace{-0.2cm}

  \captionsetup{width=0.98\textwidth}
  \caption{Effect of varying the preference centroid coefficient ($\bm{\phi}$) on PPD.\label{fig:phi_variations_recall}}
  \vspace{-0.1cm}
\end{figure*}
\begin{figure*}
  \centering
  \captionsetup{justification=raggedright, margin=0cm}
  \captionsetup[subfigure]{labelfont=normalfont, textfont=normalfont}

  \begin{subfigure}{0.28\textwidth}
    \includegraphics[width=\linewidth]{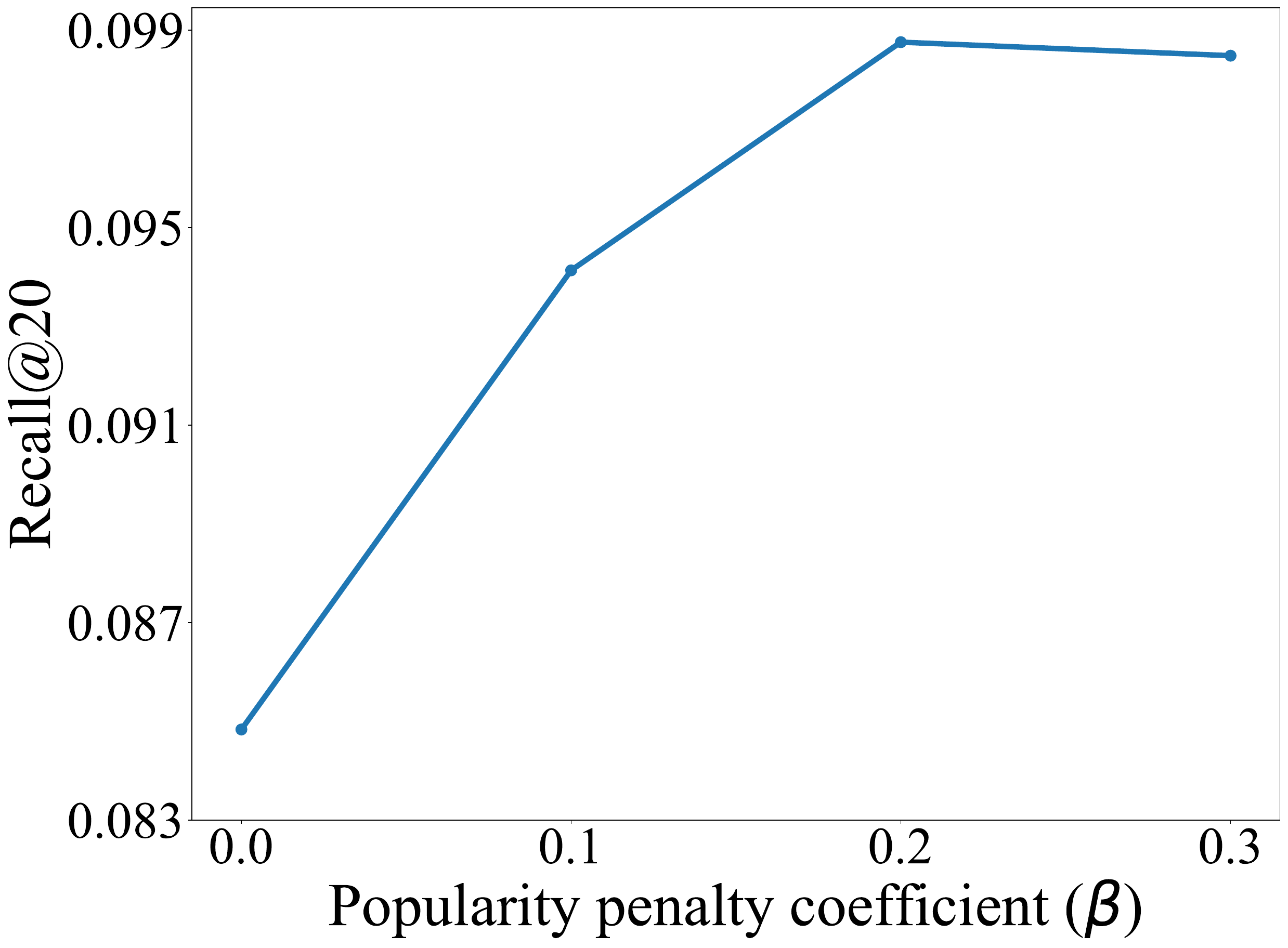}
    \caption{KuaiRec dataset}
  \end{subfigure}
  \begin{subfigure}{0.28\textwidth}
    \includegraphics[width=\linewidth]{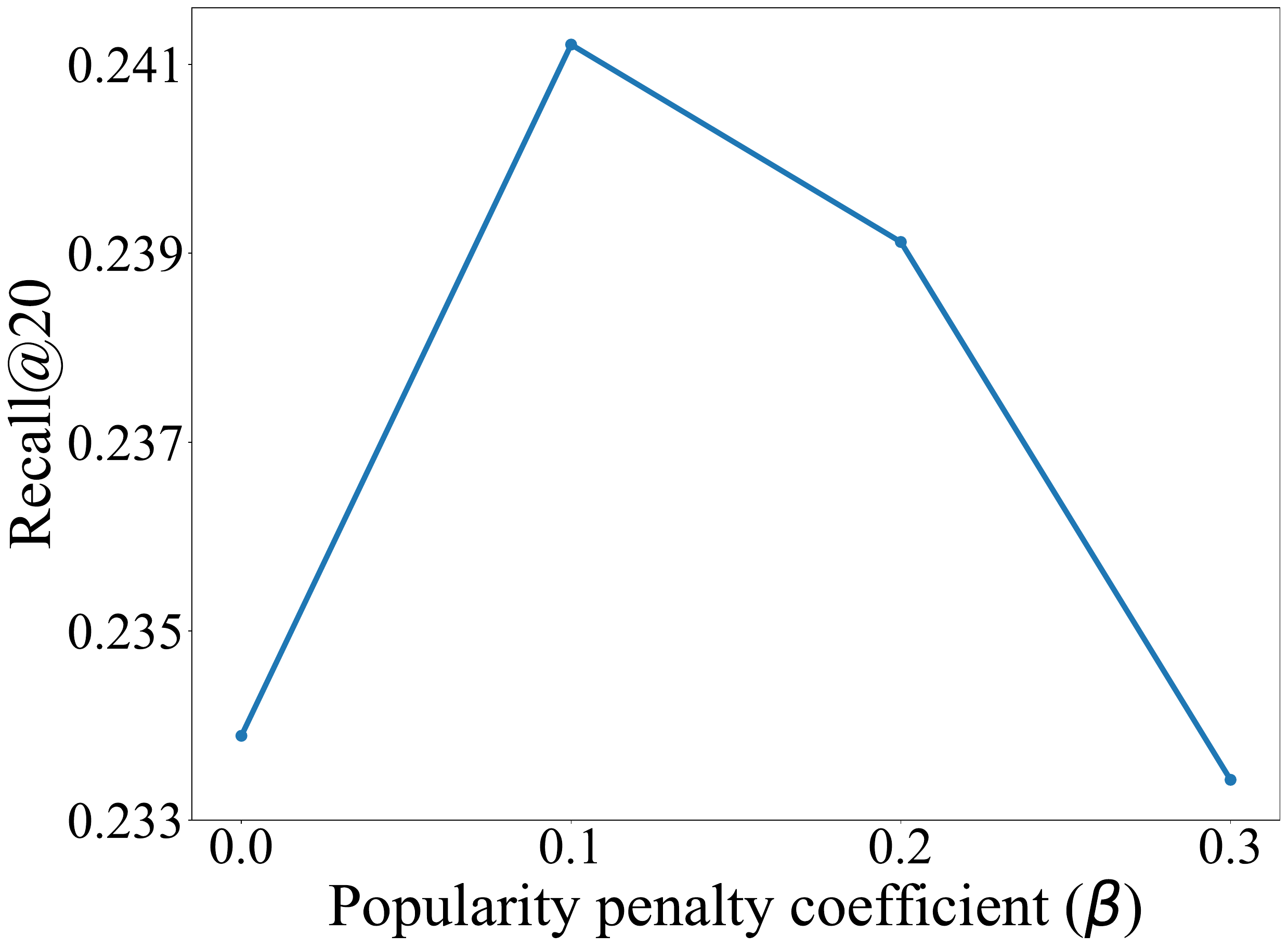}
    \caption{Coat dataset}
  \end{subfigure}
  \begin{subfigure}{0.28\textwidth}
    \includegraphics[width=\linewidth]{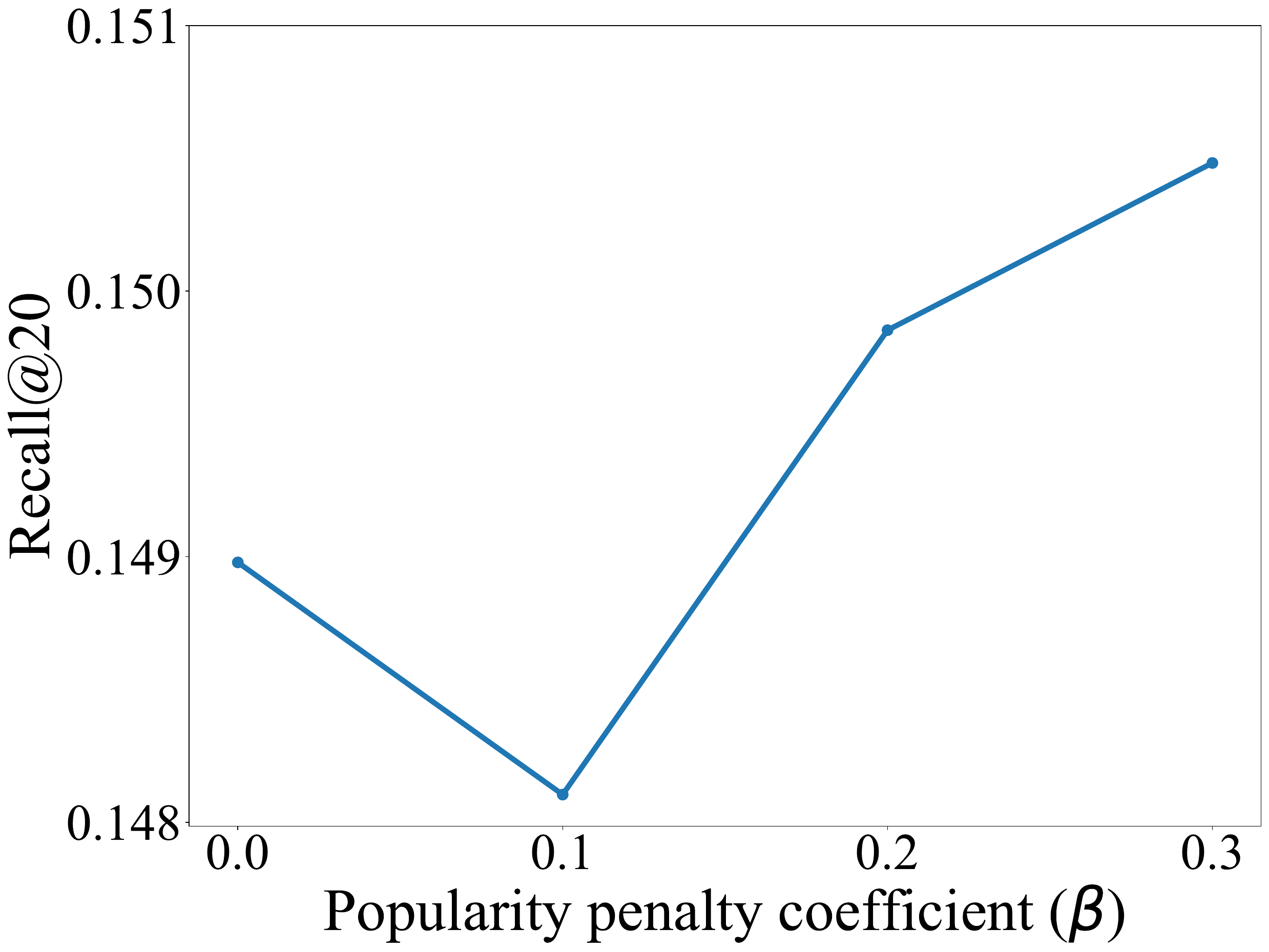}
    \caption{Yahoo! R3 dataset}
  \end{subfigure}
  \vspace{-0.2cm}

  \captionsetup{width=0.98\textwidth}
  \caption{Effect of varying the popularity penalty coefficient ($\bm{\beta}$) in personalized preference on PPD.\label{fig:beta_variations_recall}}
  \vspace{-0.15cm}
\end{figure*}

\subsection{Additional experimental results} \label{sec:add_exp}

\textbf{Performance of PPD with SGL as backbone (RQ4).}
To validate the generalizability of our method, we apply PPD to another widely used GNN-based CF backbone, Self-supervised Graph Learning (SGL)~\cite{wu-sigir21}. SGL augments LightGCN with 
self-supervised contrastive learning, generating multiple graph views via node dropout, edge dropout, and random walks. By maximizing agreement between views of the same node, SGL enhances representation robustness and recommendation accuracy. Evaluating PPD on SGL tests whether our debiasing strategy benefits other GNN-based CF backbones. Table~\ref{tab:ppd_sgl_results} shows results on KuaiRec, Coat, and Yahoo! R3. On KuaiRec, PPD achieves $+1017.4\%$ Recall@20, $+1811.1\%$ NDCG@20, and $+287.6\%$ HR@20 over SGL. For Coat, improvements are $+5.8\%$ Recall@20, $+7.3\%$ NDCG@20, and $+5.6\%$ HR@20, while on Yahoo! R3, gains are $+8.1\%$ Recall@20, $+3.1\%$ NDCG@20, and $+6.4\%$ HR@20. These results highlight two key findings. First, PPD shows substantial improvements on datasets with strong popularity bias, such as KuaiRec, where SGL alone struggles to capture unbiased signals. Second, even on less biased datasets like Coat and Yahoo! R3, PPD still yields consistent improvements. This demonstrates that PPD is not limited to a specific backbone but can be effectively applied on top of stronger backbones like SGL, further boosting recommendation accuracy through mitigating popularity bias. SGL is considered a more effective method than LightGCN. PPD improves performance across both backbones, even with the weaker LightGCN embeddings. This indicates that our popularity direction remains stable even with weaker embeddings and improves performance in both weaker and stronger recommendation models.

\textbf{Impact of incorporating personalized preference into popularity score estimation (RQ5).} In RQ5, we examine the role of personalized preference ($r_{ui}$) in estimating the popularity score ($b_{ui}$). To this end, we introduce a variant, PPD w/o-PP, which computes the popularity score solely based on global preference ($p_i$) without accounting for user-specific  preferences. The results in Table~\ref{tab:ppd_without_rui} show that PPD consistently outperforms PPD w/o-PP across all datasets and evaluation metrics. This comparison reveals that relying only on global preference can be insufficient, as popular items may still align with true user interests~\cite{zhao-kde22}. Without personalized preference, the popularity score ($b_{ui}$) risks suppressing relevant popular items, conflating popularity-driven interactions with true user preferences. This is also evident from the NAVIP~\cite{kim-cikm22} method, which relies only on global preferences (from total interactions), and PPD clearly outperforms it (see Table~\ref{tab:main_results}). 
These observations demonstrate that accounting for personalized preference in estimating the popularity score is important, as it enables PPD to more accurately estimate the true popularity effect, enabling better performance.

\textbf{Effect of the preference centroid coefficient ($\bm{\phi}$) on PPD (RQ6).}
In RQ6, we study how the preference centroid coefficient $\phi \in [0,1]$, which controls how much the preference centroid contributes when defining the popularity direction and offers flexibility in balancing preference and popularity signals, affects the performance of PPD. Figure~\ref{fig:phi_variations_recall} presents the Recall@20 results on the KuaiRec, Coat, and Yahoo! R3 datasets as $\phi$ varies from 0 to 1. On KuaiRec, performance drops at $\phi=0.25$, peaks at $\phi=0.50$, and then declines, reflecting the dataset’s severe popularity dominance and weak preference signals. When $\phi$ is too high, noisy preferences distort the estimation of the popularity direction, whereas a moderate value ($\phi=0.50$) provides the best balance, indicating that a certain level of preference signal is useful for estimating the popularity direction. For Coat, Recall@20 decreases at $\phi=0.25$ but then steadily improves, reaching its highest value at $\phi=1.0$. This pattern reflects the dataset’s milder popularity bias, which can provide more reliable preference signals and make stronger preference centroids increasingly beneficial. Yahoo! R3 shows a similar monotonic increase, achieving its best performance at $\phi=1.0$. This may happen even though preference information remains reliable under sparsity, and full reliance on it enables the most effective debiasing. Overall, these results indicate that the optimal choice of $\phi$ is dataset-dependent: moderate values are best when preference signals are weak (KuaiRec), while larger values yield stronger debiasing when preferences are reliable (Coat and Yahoo! R3).

\textbf{Effect of the popularity penalty coefficient ($\bm{\beta}$) on PPD (RQ7).}
In RQ7, we evaluate the effect of the popularity penalty coefficient ($\beta$), which controls the strength of penalizing using global preference when estimating personalized preferences. As shown in Figure~\ref{fig:beta_variations_recall}, performance varies across datasets and values of $\beta$. On KuaiRec, Recall@20 improves as $\beta$ increases up to 0.2 but declines when raised to 0.3, indicating that moderate global preference penalization yields the best personalized preference and, consequently, the highest performance. On Coat, the highest performance occurs at $\beta = 0.1$, with larger values leading to reduced Recall@20. In contrast, Yahoo! R3 achieves its peak performance at $\beta = 0.3$. These results highlight two key insights. First, without any penalty ($\beta = 0.0$), the model cannot achieve its best performance, demonstrating the necessity of incorporating a popularity penalty in estimating personalized preferences. Second, the optimal value of $\beta$ is dataset-dependent, indicating that it should be tuned for each dataset to achieve optimal performance.

%% file: main.bib
@String{Computing = "Computing" }

@String{Computer = "{IEEE} Computer" }

@String{Springer = "Springer-Verlag" }

@inproceedings{zhou-sigir23,
  title={Adaptive popularity debiasing aggregator for graph collaborative filtering},
  author={Zhou, Huachi and Chen, Hao and Dong, Junnan and Zha, Daochen and Zhou, Chuang and Huang, Xiao},
  booktitle={Proceedings of the 46th International ACM SIGIR Conference on Research and Development in Information Retrieval},
  pages={7--17},
  year={2023}
}

@inproceedings{kim-cikm22,
  title={Debiasing neighbor aggregation for graph neural network in recommender systems},
  author={Kim, Minseok and Oh, Jinoh and Do, Jaeyoung and Lee, Sungjin},
  booktitle={Proceedings of the 31st ACM International Conference on Information \& Knowledge Management},
  pages={4128--4132},
  year={2022}
}

@inproceedings{wu-kdd23,
  title={Certified edge unlearning for graph neural networks},
  author={Wu, Kun and Shen, Jie and Ning, Yue and Wang, Ting and Wang, Wendy Hui},
  booktitle={Proceedings of the 29th ACM SIGKDD Conference on Knowledge Discovery and Data Mining},
  pages={2606--2617},
  year={2023}
}

@inproceedings{gao-cikm22,
  title={KuaiRec: A fully-observed dataset and insights for evaluating recommender systems},
  author={Gao, Chongming and Li, Shijun and Lei, Wenqiang and Chen, Jiawei and Li, Biao and Jiang, Peng and He, Xiangnan and Mao, Jiaxin and Chua, Tat-Seng},
  booktitle={Proceedings of the 31st ACM International Conference on Information \& Knowledge Management},
  pages={540--550},
  year={2022}
}

@inproceedings{schnabel-icml16,
  title={Recommendations as treatments: Debiasing learning and evaluation},
  author={Schnabel, Tobias and Swaminathan, Adith and Singh, Ashudeep and Chandak, Navin and Joachims, Thorsten},
  booktitle={international conference on machine learning},
  pages={1670--1679},
  year={2016},
  organization={PMLR}
}

@inproceedings{krichene-sigkdd20,
  title={On sampled metrics for item recommendation},
  author={Krichene, Walid and Rendle, Steffen},
  booktitle={Proceedings of the 26th ACM SIGKDD international conference on knowledge discovery \& data mining},
  pages={1748--1757},
  year={2020}
}

@inproceedings{he-sigir20,
  title={Lightgcn: Simplifying and powering graph convolution network for recommendation},
  author={He, Xiangnan and Deng, Kuan and Wang, Xiang and Li, Yan and Zhang, Yongdong and Wang, Meng},
  booktitle={Proceedings of the 43rd International ACM SIGIR conference on research and development in Information Retrieval},
  pages={639--648},
  year={2020}
}

@inproceedings{joachims-wsdm17,
  title={Unbiased learning-to-rank with biased feedback},
  author={Joachims, Thorsten and Swaminathan, Adith and Schnabel, Tobias},
  booktitle={Proceedings of the tenth ACM international conference on web search and data mining},
  pages={781--789},
  year={2017}
}

@inproceedings{ning-www24,
  title={Debiasing recommendation with personal popularity},
  author={Ning, Wentao and Cheng, Reynold and Yan, Xiao and Kao, Ben and Huo, Nan and Haldar, Nur Al Hasan and Tang, Bo},
  booktitle={Proceedings of the ACM Web Conference 2024},
  pages={3400--3409},
  year={2024}
}

@article{chen-front24,
  title={How graph convolutions amplify popularity bias for recommendation?},
  author={Chen, Jiajia and Wu, Jiancan and Chen, Jiawei and Xin, Xin and Li, Yong and He, Xiangnan},
  journal={Frontiers of Computer Science},
  volume={18},
  number={5},
  pages={185603},
  year={2024},
  publisher={Springer}
}

@inproceedings{wei-sigkdd21,
  title={Model-agnostic counterfactual reasoning for eliminating popularity bias in recommender system},
  author={Wei, Tianxin and Feng, Fuli and Chen, Jiawei and Wu, Ziwei and Yi, Jinfeng and He, Xiangnan},
  booktitle={Proceedings of the 27th ACM SIGKDD conference on knowledge discovery \& data mining},
  pages={1791--1800},
  year={2021}
}

@inproceedings{gruson-wsdm19,
  title={Offline evaluation to make decisions about playlist recommendation algorithms},
  author={Gruson, Alois and Chandar, Praveen and Charbuillet, Christophe and McInerney, James and Hansen, Samantha and Tardieu, Damien and Carterette, Ben},
  booktitle={Proceedings of the Twelfth ACM International Conference on Web Search and Data Mining},
  pages={420--428},
  year={2019}
}

@article{boratto-ipm21,
  title={Connecting user and item perspectives in popularity debiasing for collaborative recommendation},
  author={Boratto, Ludovico and Fenu, Gianni and Marras, Mirko},
  journal={Information Processing \& Management},
  volume={58},
  number={1},
  pages={102387},
  year={2021},
  publisher={Elsevier}
}

@inproceedings{marlin-recsys09,
  title={Collaborative prediction and ranking with non-random missing data},
  author={Marlin, Benjamin M and Zemel, Richard S},
  booktitle={Proceedings of the third ACM conference on Recommender systems},
  pages={5--12},
  year={2009}
}

@article{su-aai09,
  title={A survey of collaborative filtering techniques},
  author={Su, Xiaoyuan and Khoshgoftaar, Taghi M},
  journal={Advances in artificial intelligence},
  volume={2009},
  number={1},
  pages={421425},
  year={2009},
  publisher={Wiley Online Library}
}

@inproceedings{he-www17,
  title={Neural collaborative filtering},
  author={He, Xiangnan and Liao, Lizi and Zhang, Hanwang and Nie, Liqiang and Hu, Xia and Chua, Tat-Seng},
  booktitle={Proceedings of the 26th international conference on world wide web},
  pages={173--182},
  year={2017}
}

@inproceedings{yang-sigir21,
  title={Enhanced graph learning for collaborative filtering via mutual information maximization},
  author={Yang, Yonghui and Wu, Le and Hong, Richang and Zhang, Kun and Wang, Meng},
  booktitle={Proceedings of the 44th international ACM SIGIR conference on research and development in information retrieval},
  pages={71--80},
  year={2021}
}

@inproceedings{gao-www22,
  title={Graph neural networks for recommender system},
  author={Gao, Chen and Wang, Xiang and He, Xiangnan and Li, Yong},
  booktitle={Proceedings of the fifteenth ACM international conference on web search and data mining},
  pages={1623--1625},
  year={2022}
}

@inproceedings{ai-sigir18,
  title={Unbiased learning to rank with unbiased propensity estimation},
  author={Ai, Qingyao and Bi, Keping and Luo, Cheng and Guo, Jiafeng and Croft, W Bruce},
  booktitle={The 41st international ACM SIGIR conference on research \& development in information retrieval},
  pages={385--394},
  year={2018}
}

@inproceedings{chen-aaai20,
  title={Measuring and relieving the over-smoothing problem for graph neural networks from the topological view},
  author={Chen, Deli and Lin, Yankai and Li, Wei and Li, Peng and Zhou, Jie and Sun, Xu},
  booktitle={Proceedings of the AAAI conference on artificial intelligence},
  volume={34},
  number={04},
  pages={3438--3445},
  year={2020}
}

@inproceedings{wang-kdd21,
  title={Deconfounded recommendation for alleviating bias amplification},
  author={Wang, Wenjie and Feng, Fuli and He, Xiangnan and Wang, Xiang and Chua, Tat-Seng},
  booktitle={Proceedings of the 27th ACM SIGKDD conference on knowledge discovery \& data mining},
  pages={1717--1725},
  year={2021}
}

@inproceedings{abdollahpouri-arXiv19,
  title={Managing popularity bias in recommender systems with personalized re-ranking},
  author={Abdollahpouri, Himan and Burke, Robin and Mobasher, Bamshad},
  booktitle={FLAIRS},
  year={2019}
}

@inproceedings{mao-cikm21,
  title={UltraGCN: ultra simplification of graph convolutional networks for recommendation},
  author={Mao, Kelong and Zhu, Jieming and Xiao, Xi and Lu, Biao and Wang, Zhaowei and He, Xiuqiang},
  booktitle={Proceedings of the 30th ACM international conference on information \& knowledge management},
  pages={1253--1262},
  year={2021}
}

@inproceedings{wang-sigir19,
  title={Neural graph collaborative filtering},
  author={Wang, Xiang and He, Xiangnan and Wang, Meng and Feng, Fuli and Chua, Tat-Seng},
  booktitle={Proceedings of the 42nd international ACM SIGIR conference on Research and development in Information Retrieval},
  pages={165--174},
  year={2019}
}

@inproceedings{kipf-iclr17,
      title={Semi-Supervised Classification with Graph Convolutional Networks}, 
      author={Thomas N. Kipf and Max Welling},
      booktitle={The Fifth International Conference on Learning Representations},
      year={2017}
}

@article{rendle-arxiv12,
  title={BPR: Bayesian personalized ranking from implicit feedback},
  author={Rendle, Steffen and Freudenthaler, Christoph and Gantner, Zeno and Schmidt-Thieme, Lars},
  journal={arXiv preprint arXiv:1205.2618},
  year={2012}
}

@inproceedings{zhang-sigir21,
  title={Causal intervention for leveraging popularity bias in recommendation},
  author={Zhang, Yang and Feng, Fuli and He, Xiangnan and Wei, Tianxin and Song, Chonggang and Ling, Guohui and Zhang, Yongdong},
  booktitle={Proceedings of the 44th international ACM SIGIR conference on research and development in information retrieval},
  pages={11--20},
  year={2021}
}

@inproceedings{oosterhuis-sigir20,
author = {Oosterhuis, Harrie and de Rijke, Maarten},
title = {Policy-Aware Unbiased Learning to Rank for Top-k Rankings},
year = {2020},
publisher = {Association for Computing Machinery},
booktitle = {Proceedings of the 43rd International ACM SIGIR Conference on Research and Development in Information Retrieval},
pages = {489–498},
numpages = {10}
}

@inproceedings{wang-sigir16,
author = {Wang, Xuanhui and Bendersky, Michael and Metzler, Donald and Najork, Marc},
title = {Learning to Rank with Selection Bias in Personal Search},
year = {2016},
publisher = {Association for Computing Machinery},
address = {New York, NY, USA},
booktitle = {Proceedings of the 39th International ACM SIGIR Conference on Research and Development in Information Retrieval},
pages = {115–124}
}

@inproceedings{luo-wsdm23,
  title={Model-based unbiased learning to rank},
  author={Luo, Dan and Zou, Lixin and Ai, Qingyao and Chen, Zhiyu and Yin, Dawei and Davison, Brian D},
  booktitle={Proceedings of the Sixteenth ACM International Conference on Web Search and Data Mining},
  pages={895--903},
  year={2023}
}

@inproceedings{luo-sigir24,
  title={Unbiased Learning-to-Rank Needs Unconfounded Propensity Estimation},
  author={Luo, Dan and Zou, Lixin and Ai, Qingyao and Chen, Zhiyu and Li, Chenliang and Yin, Dawei and Davison, Brian D},
  booktitle={Proceedings of the 47th International ACM SIGIR Conference on Research and Development in Information Retrieval},
  pages={1535--1545},
  year={2024}
}

@inproceedings{chen-sigir22,
  title={Generative adversarial framework for cold-start item recommendation},
  author={Chen, Hao and Wang, Zefan and Huang, Feiran and Huang, Xiao and Xu, Yue and Lin, Yishi and He, Peng and Li, Zhoujun},
  booktitle={Proceedings of the 45th International ACM SIGIR Conference on Research and Development in Information Retrieval},
  pages={2565--2571},
  year={2022}
}

@article{zhang-neurips22,
  title={Incorporating bias-aware margins into contrastive loss for collaborative filtering},
  author={Zhang, An and Ma, Wenchang and Wang, Xiang and Chua, Tat-Seng},
  journal={Advances in Neural Information Processing Systems},
  volume={35},
  pages={7866--7878},
  year={2022}
}

@inproceedings{chen-sigir20,
  title={Esam: Discriminative domain adaptation with non-displayed items to improve long-tail performance},
  author={Chen, Zhihong and Xiao, Rong and Li, Chenliang and Ye, Gangfeng and Sun, Haochuan and Deng, Hongbo},
  booktitle={Proceedings of the 43rd international ACM SIGIR conference on research and development in information retrieval},
  pages={579--588},
  year={2020}
}

@inproceedings{rhee-recsys22,
  title={Countering popularity bias by regularizing score differences},
  author={Rhee, Wondo and Cho, Sung Min and Suh, Bongwon},
  booktitle={Proceedings of the 16th ACM conference on recommender systems},
  pages={145--155},
  year={2022}
}

@article{bottou-jmlr13,
  title={Counterfactual reasoning and learning systems: The example of computational advertising},
  author={Bottou, L{\'e}on and Peters, Jonas and Qui{\~n}onero-Candela, Joaquin and Charles, Denis X and Chickering, D Max and Portugaly, Elon and Ray, Dipankar and Simard, Patrice and Snelson, Ed},
  journal={The Journal of Machine Learning Research},
  volume={14},
  number={1},
  pages={3207--3260},
  year={2013},
  publisher={JMLR. org}
}

@inproceedings{ovaisi-sigir21,
  title={Propensity-independent bias recovery in offline learning-to-rank systems},
  author={Ovaisi, Zohreh and Vasilaky, Kathryn and Zheleva, Elena},
  booktitle={Proceedings of the 44th International ACM SIGIR Conference on Research and Development in Information Retrieval},
  pages={1763--1767},
  year={2021}
}

@inproceedings{wasilewski-flairs16,
  title={Incorporating Diversity in a Learning to Rank Recommender System.},
  author={Wasilewski, Jacek and Hurley, Neil},
  booktitle={FLAIRS},
  pages={572--578},
  year={2016}
}

@article{kamishima-recsys14,
  title={Correcting popularity bias by enhancing recommendation neutrality.},
  author={Kamishima, Toshihiro and Akaho, Shotaro and Asoh, Hideki and Sakuma, Jun},
  journal={RecSys posters},
  volume={10},
  year={2014}
}

@inproceedings{bonner-recsys18,
  title={Causal embeddings for recommendation},
  author={Bonner, Stephen and Vasile, Flavian},
  booktitle={Proceedings of the 12th ACM conference on recommender systems},
  pages={104--112},
  year={2018}
}

@inproceedings{he-icds22,
  title={Mitigating popularity bias in recommendation via counterfactual inference},
  author={He, Ming and Li, Changshu and Hu, Xinlei and Chen, Xin and Wang, Jiwen},
  booktitle={International Conference on Database Systems for Advanced Applications},
  pages={377--388},
  year={2022},
  organization={Springer}
}

@article{zhao-kde22,
  title={Popularity bias is not always evil: Disentangling benign and harmful bias for recommendation},
  author={Zhao, Zihao and Chen, Jiawei and Zhou, Sheng and He, Xiangnan and Cao, Xuezhi and Zhang, Fuzheng and Wu, Wei},
  journal={IEEE Transactions on Knowledge and Data Engineering},
  volume={35},
  number={10},
  pages={9920--9931},
  year={2022},
  publisher={IEEE}
}

@inproceedings{zhao-www25,
  title={Distributionally robust graph out-of-distribution recommendation via diffusion model},
  author={Zhao, Chu and Yang, Enneng and Liang, Yuliang and Zhao, Jianzhe and Guo, Guibing and Wang, Xingwei},
  booktitle={Proceedings of the ACM on Web Conference 2025},
  pages={2018--2031},
  year={2025}
}

@inproceedings{zha-kdd22,
  title={Autoshard: Automated embedding table sharding for recommender systems},
  author={Zha, Daochen and Feng, Louis and Bhushanam, Bhargav and Choudhary, Dhruv and Nie, Jade and Tian, Yuandong and Chae, Jay and Ma, Yinbin and Kejariwal, Arun and Hu, Xia},
  booktitle={Proceedings of the 28th ACM SIGKDD Conference on Knowledge Discovery and Data Mining},
  pages={4461--4471},
  year={2022}
}

@inproceedings{chen-cikm20,
  title={Label-aware graph convolutional networks},
  author={Chen, Hao and Xu, Yue and Huang, Feiran and Deng, Zengde and Huang, Wenbing and Wang, Senzhang and He, Peng and Li, Zhoujun},
  booktitle={Proceedings of the 29th ACM international conference on information \& knowledge management},
  pages={1977--1980},
  year={2020}
}

@inproceedings{dong-www23,
  title={Hierarchy-aware multi-hop question answering over knowledge graphs},
  author={Dong, Junnan and Zhang, Qinggang and Huang, Xiao and Duan, Keyu and Tan, Qiaoyu and Jiang, Zhimeng},
  booktitle={Proceedings of the ACM web conference 2023},
  pages={2519--2527},
  year={2023}
}

@inproceedings{huang-icml22,
  title={Going deeper into permutation-sensitive graph neural networks},
  author={Huang, Zhongyu and Wang, Yingheng and Li, Chaozhuo and He, Huiguang},
  booktitle={International conference on machine learning},
  pages={9377--9409},
  year={2022},
  organization={PMLR}
}

@inproceedings{zhang-wsdm23,
  title={Efficiently leveraging multi-level user intent for session-based recommendation via atten-mixer network},
  author={Zhang, Peiyan and Guo, Jiayan and Li, Chaozhuo and Xie, Yueqi and Kim, Jae Boum and Zhang, Yan and Xie, Xing and Wang, Haohan and Kim, Sunghun},
  booktitle={Proceedings of the sixteenth ACM international conference on web search and data mining},
  pages={168--176},
  year={2023}
}

@inproceedings{zhang-sigir22,
  title={Geometric disentangled collaborative filtering},
  author={Zhang, Yiding and Li, Chaozhuo and Xie, Xing and Wang, Xiao and Shi, Chuan and Liu, Yuming and Sun, Hao and Zhang, Liangjie and Deng, Weiwei and Zhang, Qi},
  booktitle={Proceedings of the 45th international ACM SIGIR conference on research and development in information retrieval},
  pages={80--90},
  year={2022}
}

@inproceedings{sun-sigir2020,
  title={Neighbor interaction aware graph convolution networks for recommendation},
  author={Sun, Jianing and Zhang, Yingxue and Guo, Wei and Guo, Huifeng and Tang, Ruiming and He, Xiuqiang and Ma, Chen and Coates, Mark},
  booktitle={Proceedings of the 43rd international ACM SIGIR conference on research and development in information retrieval},
  pages={1289--1298},
  year={2020}
}

@inproceedings{wu-sigir21,
  title={Self-supervised graph learning for recommendation},
  author={Wu, Jiancan and Wang, Xiang and Feng, Fuli and He, Xiangnan and Chen, Liang and Lian, Jianxun and Xie, Xing},
  booktitle={Proceedings of the 44th international ACM SIGIR conference on research and development in information retrieval},
  pages={726--735},
  year={2021}
}

@inproceedings{lin-www22,
  title={Improving graph collaborative filtering with neighborhood-enriched contrastive learning},
  author={Lin, Zihan and Tian, Changxin and Hou, Yupeng and Zhao, Wayne Xin},
  booktitle={Proceedings of the ACM web conference 2022},
  pages={2320--2329},
  year={2022}
}

@inproceedings{wang-kdd22,
  title={Towards representation alignment and uniformity in collaborative filtering},
  author={Wang, Chenyang and Yu, Yuanqing and Ma, Weizhi and Zhang, Min and Chen, Chong and Liu, Yiqun and Ma, Shaoping},
  booktitle={Proceedings of the 28th ACM SIGKDD conference on knowledge discovery and data mining},
  pages={1816--1825},
  year={2022}
}

@inproceedings{yu-sigir22,
  title={Are graph augmentations necessary? simple graph contrastive learning for recommendation},
  author={Yu, Junliang and Yin, Hongzhi and Xia, Xin and Chen, Tong and Cui, Lizhen and Nguyen, Quoc Viet Hung},
  booktitle={Proceedings of the 45th international ACM SIGIR conference on research and development in information retrieval},
  pages={1294--1303},
  year={2022}
}

@article{zhang-neurips23,
  title={Empowering collaborative filtering with principled adversarial contrastive loss},
  author={Zhang, An and Sheng, Leheng and Cai, Zhibo and Wang, Xiang and Chua, Tat-Seng},
  journal={Advances in Neural Information Processing Systems},
  volume={36},
  pages={6242--6266},
  year={2023}
}

@inproceedings{zhu-kdd21,
  title={Popularity bias in dynamic recommendation},
  author={Zhu, Ziwei and He, Yun and Zhao, Xing and Caverlee, James},
  booktitle={Proceedings of the 27th ACM SIGKDD conference on knowledge discovery \& data mining},
  pages={2439--2449},
  year={2021}
}

@inproceedings{steck-recsys18,
  title={Calibrated recommendations},
  author={Steck, Harald},
  booktitle={Proceedings of the 12th ACM conference on recommender systems},
  pages={154--162},
  year={2018}
}

@article{yu-kde23,
  title={XSimGCL: Towards extremely simple graph contrastive learning for recommendation},
  author={Yu, Junliang and Xia, Xin and Chen, Tong and Cui, Lizhen and Hung, Nguyen Quoc Viet and Yin, Hongzhi},
  journal={IEEE Transactions on Knowledge and Data Engineering},
  volume={36},
  number={2},
  pages={913--926},
  year={2023},
  publisher={IEEE}
}

@inproceedings{chaney-recsys18,
  title={How algorithmic confounding in recommendation systems increases homogeneity and decreases utility},
  author={Chaney, Allison JB and Stewart, Brandon M and Engelhardt, Barbara E},
  booktitle={Proceedings of the 12th ACM conference on recommender systems},
  pages={224--232},
  year={2018}
}

@inproceedings{canamares-sigir18,
  title={Should I follow the crowd? A probabilistic analysis of the effectiveness of popularity in recommender systems},
  author={Ca{\~n}amares, Roc{\'\i}o and Castells, Pablo},
  booktitle={The 41st International ACM SIGIR Conference on Research \& Development in Information Retrieval},
  pages={415--424},
  year={2018}
}

@article{yao-neurips17,
  title={Beyond parity: Fairness objectives for collaborative filtering},
  author={Yao, Sirui and Huang, Bert},
  journal={Advances in neural information processing systems},
  volume={30},
  year={2017}
}

@inproceedings{wang-sigir20,
  title={Disentangled graph collaborative filtering},
  author={Wang, Xiang and Jin, Hongye and Zhang, An and He, Xiangnan and Xu, Tong and Chua, Tat-Seng},
  booktitle={Proceedings of the 43rd international ACM SIGIR conference on research and development in information retrieval},
  pages={1001--1010},
  year={2020}
}

@book{perwass-springer09,
  title={Geometric algebra with applications in engineering},
  author={Perwass, Christian},
  year={2009},
  publisher={Springer}
}

@inproceedings{zhang-kdd23,
  title={Empowering long-tail item recommendation through cross decoupling network (cdn)},
  author={Zhang, Yin and Wang, Ruoxi and Cheng, Derek Zhiyuan and Yao, Tiansheng and Yi, Xinyang and Hong, Lichan and Caverlee, James and Chi, Ed H},
  booktitle={Proceedings of the 29th ACM SIGKDD Conference on Knowledge Discovery and Data Mining},
  pages={5608--5617},
  year={2023}
}
